\documentclass[12pt,preprint]{aastex}

\newcommand{\Hb}{\ensuremath{{\rm H}\beta}}
\newcommand{\Mgb}{\ensuremath{{\rm Mg}\, b}}
\newcommand{\Fe}{\ensuremath{\langle {\rm Fe}\rangle}}
\newcommand{\MgFe}{\ensuremath{[{\rm MgFe}]^{\prime}}}
\newcommand{\aFe}{\ensuremath{\alpha/{\rm Fe}}}

\newcommand{\ZH}{\ensuremath{Z/{\rm H}}}

\begin{document}
\title{Oxygen and neon abundances of planetary nebulae in
the elliptical galaxy NGC 4697\footnote{
Part of the data presented herein were obtained at the European 
Southern observatory, Chile, Program ESO 69.B-0124. 
Part of the data presented herein were obtained at the W.M. Keck 
Observatory, which 
is operated as a scientific partnership among the California Institute of 
Technology, the University of California and the National Aeronautics and 
Space Administration. The Observatory was made possible by the generous 
financial support of the W.M. Keck Foundation.}}
\author{R. H. M\'endez\altaffilmark{2}, D. Thomas\altaffilmark{3},
        R. P. Saglia\altaffilmark{4}, C. Maraston\altaffilmark{3},
        R. P. Kudritzki\altaffilmark{2}, and R. Bender\altaffilmark{4,5}}
\altaffiltext{2}{Institute for Astronomy, University of Hawaii, 
                 2680 Woodlawn Drive, Honolulu, HI 96822}
\email{mendez@ifa.hawaii.edu}
\altaffiltext{3}{Max-Planck-Institut f\"ur extraterrestrische Physik,
                 Giessenbachstra\ss e, D-85748 Garching, Germany;
                 current address: University of Oxford, Denys Wilkinson
                 Building, Keble Road, Oxford OX1 3RH, UK}
\altaffiltext{4}{Max-Planck-Institut f\"ur extraterrestrische Physik,
                 Giessenbachstra\ss e, D-85748 Garching, Germany}
\altaffiltext{5}{Universit\"ats-Sternwarte M\"unchen, Scheinerstr. 1, 
                 D-81679 M\"unchen, Germany}

\begin{abstract}
This paper presents deep spectra of 14 planetary nebulae (PNs) at a variety
of angular distances from the center of the flattened elliptical galaxy
NGC 4697, which is located at $\sim$ 11 Mpc from us. Both near the center
and among the most outlying 
PNs we have found several examples of very strong [O III] $\lambda$5007, 
about 20 times brighter than H$\beta$. This, together with strong
[Ne III] lines, implies a lower limit for O 
and Ne abundances near solar at the center and also at more than two 
effective radii from the center of NGC 4697. Thus we have found, for the 
first time from individual stars, direct evidence of the existence of 
a metal-rich population in elliptical galaxies, confirming metallicities 
derived from absorption-line indices and extending the
information to angular distances where the stellar surface brightness
is too low for absorption-line studies.
A comparison with abundances obtained closer to the center of NGC 4697
from integrated absorption-line spectra indicates that the metallicity 
gradient, which is quite steep within one effective radius, 
becomes much less pronounced beyond one effective radius.
The only alternative to this interpretation would be
the presence of a very metal-poor PN population with $[\ZH]$ below $-1$,
which we cannot rule out but would require an extremely bimodal
metallicity distribution with almost no intermediate metallicities.
\end{abstract}


\section{Introduction}

How were elliptical galaxies formed, and when did that happen, are matters 
of intense contemporary discussion. The dominant idea is the hierarchical 
merging of small galaxies into larger ones, frequently pictured as a long 
process extending down to small redshifts (White \& Rees 1978). 

In this context, metallicity gradients in elliptical galaxies are important 
diagnostics. A monolithic dissipative collapse produces steep gradients
(Carlberg 1984), and mergers dilute them (White 1980). 
Observations show that metallicity gradients are present in early-type 
galaxies (Davies, Sadler \& Peletier 1993; Carollo \& Danziger 1994;
Fisher, Franx, \& Illingworth 1995; Saglia et al. 2000; Mehlert et
al. 2003; Wu et al. 2004), but they appear somewhat flatter than
predicted by the pure dissipational collapse (Carlberg 1984), which
suggests that both dissipative processes and mergers play a role.
Kobayashi (2004) gives a recent review on the 
subject, concluding that the observed variety of metallicity gradients
requires different individual merging histories, with steeper gradients
for galaxies that have not undergone major mergers.
Simulations of monolithic collapse models taking into account different
escape velocities at different galactocentric radii
produce results that are consistent with observed absorption line index
gradients inside $\sim 1$ effective radius (Martinelli et al. 1998,
Angeletti \& Giannone 2003). But they predict a significant steepening
beyond that limit, which still needs to be checked observationally.

On the other hand, how reliable are the observed metallicity gradients? 
They are based on colors and absorption-line strengths (Mg, Fe, H) 
measured on long-slit, integrated-light 
spectra of the galaxies. Their interpretation in terms of metallicity is 
affected by the well-known age-metallicity degeneracy. The introduction of
Balmer lines to break the degeneracy does not necessarily solve the problem, 
because strong Balmer lines can be produced not only by young stars, but also 
by blue horizontal-branch stars (de Freitas Pacheco \& Barbuy 1995; 
Maraston \& Thomas 2000; Lee, Yoon, \& Lee 2000).
There is also the complication introduced by possible variations in the 
$\alpha$/Fe abundance ratio, which may affect the derived gradients 
(Trager et al. 2000a; Proctor \& Sansom 2002; Thomas, 
Maraston \& Bender 2003). All these obstacles introduce uncertainties in
the derived ages, metallicities and metallicity gradients.
It would be desirable to determine metallicities of individual stars in
elliptical galaxies. Single stars, however, are too faint and cannot be 
resolved with current technology, with the only exception of the nearby
NGC 5128 (e.g. Harris \& Harris 2002).

Planetary nebulae (PNs) in elliptical galaxies offer a possible way to
obtain alternative information: we can detect them individually because 
of their characteristic emission-line spectra. By measuring the fluxes 
of adequate diagnostic emission lines, it would be possible to derive 
at least O and Ne abundances of PNs at different projected radii to 
directly study the metallicity gradient. In this way we would be 
able to circumvent the problem of the age-metallicity degeneracy. 
Forestini \& Charbonnel (1997) explain that no significant change in 
surface O, Ne 
abundances is expected from Asymptotic Giant Branch nucleosynthesis.
An additional important advantage of measuring PN element 
abundances is the ability to cover a considerably 
larger range in projected radii. Because of the dramatic
drop in surface brightness with increasing radius, the
derivation of metallicities from the diffuse light extends to at most
1.5 effective radii $r_{\rm eff}$, whereas PNs can be easily
detected out to $3-4~r_{\rm eff}$.

There is one problem, however: the best diagnostic lines, like [O III]
$\lambda$4363 or [S II] $\lambda\lambda$6717, 6731, are very faint.
One pioneering attempt to study a few PNs in NGC 5128 with the ESO 3.6m 
telescope was not fully successful, in the sense that no accurate estimates 
of electron temperatures and electron densities could be obtained (Walsh et 
al. 1999). In addition, the fact that NGC 5128 has suffered a recent merger
with a low-metallicity galaxy complicated the interpretation of the few 
PN abundance estimates obtained by those authors.

This kind of project clearly requires the light-collecting power of 
(at least) 8-10m telescopes. 
Having decided to explore the PN approach to abundances in 
ellipticals, we chose to avoid the very tempting NGC 5128 and selected
instead NGC 4697, a flattened elliptical at a distance of about 11 Mpc.
Closer to us than the Virgo or Fornax ellipticals, with 
$M_{\rm B} = -20.1$, NGC 4697 is of intermediate size and luminosity. 
It belongs to a loose group of 18 galaxies (Garc\'\i a 1993). 
It shows a relaxed structure but has
a small dust ring near the core that may be evidence of a past merger 
(Pinkney et al. 2003). Earlier metallicity determinations 
(Kobayashi \& Arimoto 1999; Trager et al. 2000b; 
Thomas et al. 2005) indicate a central [$Z$/H] 
slightly above solar, and a relatively steep negative gradient.

Our choice of NGC 4697 was strongly influenced by the fact that we had
already found more than 500 PNs in this galaxy (M\'endez et al. 2001).
Knowing the fluxes of the brightest PNs in [O III] $\lambda$5007 we 
could estimate that we had a realistic chance of detecting [O III] 
$\lambda$4363, which is a factor 100 weaker than $\lambda$5007 for PN 
electron temperatures of about 12,000 K.

This paper is organized in the following way: in Section 2 we present
new observations of absorption-line strengths from long-slit spectra
of NGC 4697, analyzed with modern techniques to try to refine our 
knowledge of the metallicity gradient determined in the classical way.
Then we introduce the PN observations.
Section 3 describes the PN selection in NGC 4697 from our previous work,
and the VLT FORS observations and reductions of PN spectra. Section 4
describes in a similar way the complementary Keck I LRIS-B spectra
of the same PNs. Section 5 describes the resulting line intensities and
section 6 the analysis technique. It turns out that we could not detect 
[O III] $\lambda$4363 in the individual PN spectra, but somewhat 
accidentally we can infer interesting limits
on the O and Ne abundances in PNs at different angular distances from the 
center of NGC 4697. Section 7 presents a discussion of the results and 
what constraints can be derived from them. 
Section 8 provides a summary of the conclusions 
and some perspectives for future work.

\section{Absorption-line studies: long-slit spectroscopy}

\subsection{Observations}

We obtained long-slit spectroscopy along the major and minor axes of 
NGC 4697 using the second Focal Reducer and Spectrograph (FORS2) at the 
Cassegrain focus of Unit Telescope 2 (UT2) of the ESO Very Large Telescope,
Cerro Paranal, Chile, on the nights of 2001 February 27/28 and 
Feb 28 / March 1, under cloudy sky conditions, with 1-1.5 arcsec seeing.

FORS2 with the standard collimator gives a 
field of 6.8$\times$6.8 arc min on a 2080$\times$2048 CCD (pixel size 
24$\times$24 $\mu$m). The image scale is 0.2 arc sec pix$^{-1}$. The
observing mode was long-slit spectroscopy using grism 1400V+18. This 
grism gives a spectral resolution of 2.5 \AA, 
equivalent to $\sigma_{\rm instr}=63$ km s$^{-1}$, 
for slits 1 arcsec wide (see the FORS Manual, version 2.0). 
The spectral coverage was from 4770 to 5700 \AA. We integrated
a total of 7228 sec along the major axis and 3600 sec along the minor
axis of NGC 4697. In addition, we acquired spectra of the star
HR 3428 at several positions along the slit to be used as kinematic 
templates. 

The data reduction was made using MIDAS and followed the usual
steps: after bias subtraction and flatfielding with dome flats, cosmic
rays were eliminated using median filtering. The wavelenght
calibration was based on HeArHgCd lamps and gave 0.2 \AA\ rms residuals. 
After rebinning on a natural logarithmic wavelength scale, the sky measured
at the ends of the slit was subtracted and the galaxy spectra were
rebinned along the spatial direction to obtain nearly constant
signal-to-noise ratio with radius.

We postpone a discussion of the kinematics to a future paper, where we plan 
to combine the long-slit spectral information with the PN kinematics.

\subsection{Absorption line indices}

We have measured the Lick absorption line indices \Hb, \Mgb, Fe5270,
and Fe5335 following the definitions of Worthey et al. (1994).
The spectra have been smoothed to Lick spectral resolution
($\sim 8\;$\AA\ FWHM). Corrections for the velocity dispersion broadening
have been applied, as described in Mehlert et al. (2000). Errors were
derived from Monte Carlo simulations, taking into account
the Poisson noise and the errors in the kinematics.
Within the central aperture of 1/8 of the
effective radius our measurements agree very well with the
measurements of Gonz\'alez (1993). Our measurements and derived quantities
are listed in Tables 1 and 2 (major and minor axes, respectively).

Fig. 1 shows the strengths of the line indices \Mgb, 
$\Fe=0.5 ({\rm Fe5270}+{\rm Fe5335})$, and \Hb, as a function of 
distance from the center of NGC 4697, along the major and minor axes.

We detect weak [O III] emission at
several radii (particularly at $R\approx +15$ to $+30$ arcsec along
the major axis, where H$\beta\approx 1.5$ \AA), but we do not attempt
any correction. In the central region, where a kinematically distinct 
disk is dominant, there is a sharp decrease of H$\beta$, together with 
a sharp increase of \Mgb\ and \Fe.

Fig. 2 shows the absorption line index measurements in 
a different way. We introduce 
$\MgFe\ = \sqrt{\Mgb (0.72 \times {\rm Fe5270}+0.28 \times {\rm Fe5335})}$.
The index \MgFe, a slight modification of [MgFe] defined by Gonz\'alez 
(1993), is a good tracer of total
metallicity (still degenerate with age, of course), as it is almost
completely independent of \aFe\ ratio variations (Thomas, Maraston, \& 
Bender 2003, hereafter TMB03).
We have plotted \MgFe\ vs.\ \Hb\ (left panel) and \Mgb\ vs.\ \Fe\ 
(right panel). Overplotted are the stellar population models of
TMB03 for various ages, metallicities, and \aFe\ ratios as indicated
by the labels in the diagrams. Since also \Hb\
is almost insensitive to \aFe\ (TMB03), the left panel of
Fig. 2 is well suited to reading off ages and total metallicities
independent of abundance ratio effects.  The right panel of Fig. 2
provides rough estimates for the \aFe\ ratios. The different 
symbols in Fig. 2 correspond to quantities along the major and minor 
axes. The position of the 
galaxy's center is indicated by the arrow. Measurement errors do not
exceed the symbol sizes. The \MgFe-\Hb\ plot shows that the average
($V$-luminosity weighted) ages of the stellar populations in NGC~4697
scatter about $10\;$Gyr with no significant dependence on galaxy
radius, while metallicities clearly vary between about 2 and 1/2
solar. The \aFe\ ratios are about 1.5 solar with a slight tendency to
higher values in the outer parts of the galaxy. The two radial
directions are consistent. The apparently older ages above
$15\;$Gyr at large positive radii (diamonds) are most likely not real,
and probably an artifact caused by emission-line filling of the \Hb\
absorption line.

\subsection{Stellar population parameters}

The explicit gradients are shown in Fig. 3, in which the stellar
population parameters age, total metallicity, and \aFe\ ratio are
plotted as functions of galactic radius (see also Table 1). The former 
are derived from the indices of Fig. 2 in a twofold iterative
procedure. First, we arbitrarily fix the \aFe\ ratio, and determine
ages and metallicities for the index pairs (\Hb,\Mgb) and (\Hb,\Fe),
by starting with arbitrary age-metallicity pairs, which we modify
iteratively until both index pairs are reproduced. The two
metallicities obtained from \Mgb\ and \Fe, respectively, are used to
adjust the \aFe\ ratio, and to start a new iteration. These steps are
repeated until the age-metallicity pairs derived from (\Hb,\Mgb) and
(\Hb,\Fe) at a given \aFe\ ratio are consistent within 1 per cent
accuracy. For ages and metallicities between the grid points quoted
above, we interpolate linearly.

Fig. 3 shows only a very small radial gradient in age, with ages 
around $9\;$Gyr in the center (in good agreement with Kobayashi \& 
Arimoto 1999 and Trager et al. 2000a, 2000b), and $10\;$Gyr at
larger radii. Also the \aFe\ ratio does not vary significantly with
galactic radius, hence NGC 4697 appears globally \aFe\ enhanced by
about a factor 1.6.  This lack of significant gradients in both age
and \aFe\ ratio agrees very well with a recent study by Mehlert et
al. (2003), in which on average no age and no \aFe\ ratio gradients
have been found for a sample of 35 early-type galaxies in the Coma
cluster. This lack of age gradients in early-type galaxies gets
further support from Wu et al. (2004), who obtained
the same result analyzing color gradients of 36 early-type galaxies
drawn from the Sloan Digital Sky Survey.

More interesting for the aim of this paper is the metallicity
gradient. We obtain central metallicities of about 1.5 solar
($[\ZH]\sim 0.2$) at $\log r\sim 0.5$ and a relatively steep
gradient. As a consequence, the average metallicity of the stellar
light in NGC 4697 around the effective radius ($\log r\sim 1.5$) at
about 100 arcseconds distance from the center is about 2/3 solar
($[\ZH]\sim -0.2$). Extrapolating the average gradient to 2.5 effective
radii we would predict $[\ZH]\sim -0.5$ or $-0.6$. But the data 
hint of a saturation at about one effective radius, and 
extrapolation to higher radii is certainly risky. We
conclude that the average stellar metallicity beyond one
effective radius, derived from the integrated absorption-line spectra, 
is at least a factor 1.5 below solar. In the 
following sections we investigate to what extent these predictions
are confirmed by the PN abundances.

\section{VLT observations and reductions of PNs in NGC 4697}

Reliable PN abundance determinations require detection of rather weak
diagnostic lines like [O III] $\lambda$4363. In our attempt to detect
such faint emission lines, we selected the brightest 
of the several hundred PNs in NGC 4697 (M\'endez et al. 2001) 
that could be found at different angular distances from the core of the 
galaxy along a narrow band in the sky so as to ensure adequate spectral 
coverage for all the targets in multi-slit spectroscopy.

Table 3 provides the coordinates of the 14 selected bright PNs, 
the angular distance of each PN from the core of the galaxy, as defined by 
the center of light, and the [O III] $\lambda$5007 fluxes expressed
in magnitudes $m$(5007), as measured by M\'endez et al. (2001). Finding 
charts for the 14 PNs are given in the Appendix. Fig. 4 shows schematically
the positions of the 14 PNs relative to the core of NGC 4697. 
The size and orientation of Fig. 4 imitate Fig. 7 in M\'endez et al. 
(2001). The effective radius $r_{\rm eff}$ of NGC 4697
is 95 arc seconds (Binney, Davies \& Illingworth 1990).

The VLT observations were made with the first Focal Reducer and low-dispersion
Spectrograph (FORS1) at the Cassegrain focus of Unit Telescope 3 (Melipal)
of the ESO Very Large Telescope, Cerro Paranal, Chile, on the nights of 2002
April 15/16, 16/17 and 17/18. FORS1, with the standard collimator, gives a 
field of 6.8$\times$6.8 arc min on a 2080$\times$2048 CCD (pixel size 
24$\times$24 $\mu$m). The image scale is 0.2 arc sec pix$^{-1}$. The
observing mode was multi-object spectroscopy (MOS) using grism 300V and
filter gg375. This grism gives a spectral resolution of about 10 \AA \ at
5000 \AA, for slits 1 arc sec wide. 

The slit positions for the selected PNs were determined from 
direct inspection of the discovery image, 
a combination of narrow-band exposures taken with FORS1. This
spectrograph allows to define 19 slits per exposure; 14 slits were given
to our PN targets, and the rest were assigned to faint foreground stars
to check the slit positioning (which was done by taking short exposures
without grism through the slits) and to help locate the dispersion lines
as a function of position across the field (which is important in the case
of spectra consisting of isolated emission lines).

All three nights were of photometric quality, with seeing of about 1.1
arc sec on the first night, 0.6 and 0.8 arc sec on the second and third.
We were able to complete 23 MOS exposures, each with an exposure time of
3000 seconds, all of them at air masses smaller than 1.6. Two additional
exposures during the third night were stopped at exposure times of about 
1500 seconds (tracking problems), and 1800 seconds (near the end of the 
night, because the air mass was becoming $>2$). We also took many spectra 
of the spectrophotometric standard stars LTT 3218, LTT 7379, CD-32 9927
(Hamuy et al. 1992), with typical exposure times between 10 and 30 seconds.

Note that this spectrograph is used in combination with an atmospheric
dispersion compensator in front of the Cassegrain focus. 

The reductions were made using IRAF\footnote{IRAF is distributed by the
National Optical Astronomical Observatories, operated by the Association 
of Universities for Research in Astronomy, Inc., under contract to the 
National Science Foundation} standard tasks. After bias subtraction we
carefully built a normalized flat field (IRAF tasks ``response'', ``illum'' 
applied separately to each of the 19 individual slits), 
where all pixels without a useful lamp signal were set equal to 1.
All the science exposures were divided by this normalized flat field.

Given the mechanical stability of FORS1 we decided to
register (IRAF task ``imshift'') the science exposures, using
the strong [O III] $\lambda\lambda$4959, 5007 emissions, visible in
all individual exposures, as our reference points. The images were
then combined, assigning weights proportional to the strength of 
[O III] $\lambda$5007. The image combination helped to eliminate 
cosmic ray events, and the increased signal allowed an easier 
control of the spectrum extractions.

In spectroscopy of extremely faint emission-line objects the most delicate
step is the background modeling and subtraction. Especially in our case,
because the galaxy stellar light contributes a non-uniform background.
Fortunately NGC 4697 shows a rather smooth surface brightness distribution,
which makes background modeling easier. The only complication was the 
occasional foreground star falling partly or completely on the slit. 
Even in such cases the rather long slits provided by FORS1 (20 arc seconds) 
made it possible to model the continuum and subtract it using the IRAF task 
``apall''. In some particularly difficult cases, close to the galaxy core, 
we tried to use the long-slit task ``background'' for an alternative 
background subtraction, but the results were not significantly different 
from what is produced by ``apall''. 

The wavelength calibration was done using He-Ar-Hg comparison spectra.
Standard star spectra were used for the flux calibration. Since we had
combined all the science exposures, we used the following procedure: first,
we took an individual MOS science exposure, taken in very good seeing at 
the same air mass as a standard spectrum, and compared the flux of the 
standard star with the flux of the control stars in some of the 19 FORS1 
slits, used to check the slit positioning. This flux calibration, being 
performed at the same air mass, needs no atmospheric extinction correction. 

Second, we compared the flux of the ``control stars'' in the individual
science exposure with the corresponding flux in the combined science 
exposure; this comparison provides a correction factor as a function of 
wavelength, which takes care of atmospheric extinction and also of possible
wavelength-dependent slit losses. The correction factors derived from 
control stars in different slits across the CCD (specifically slits 5, 
14 and 16) were similar, and therefore we averaged them, to obtain a final 
correction factor which we applied to all the PN slits. Having obtained
the flux calibrated spectra, the emission lines were shifted to the 
rest wavelengths and the final fluxes in erg cm$^{-2}$ s$^{-1}$ were 
obtained by integrating the emission lines over wavelength.

Initial inspection of the final reduced spectra showed many PNs with
[O III] $\lambda$5007 almost 20 times as strong as H$\beta$. This is
rather unusual in nearby galaxies with lower-than-solar metallicities
like the LMC and SMC; see e.g. the histogram in Fig. 3 of M\'endez \& 
Soffner (1997). The only similar case with high frequency of very strong 
$\lambda$5007 appears to be the sample observed by Richer, Stasinska \& 
McCall (1999) in the metal-rich bulge of M 31. 

We did not get any convincing detection of [O III] $\lambda$4363 on 
individual spectra. The UV response of FORS1 was not enough for reliable 
measurements of [Ne III]$\lambda\lambda$3868,
3967. And of course we could not detect [O II] $\lambda$3727. For all
these reasons we judged convenient to complement the FORS1 spectra with 
spectra taken with the new blue camera of LRIS, the Keck low-resolution 
imaging spectrometer. These spectra are described in next sectiom.
 
\section{Keck observations and reductions of PNs in NGC 4697}

Knowing the coordinates of the PNs, we used the software ``autoslit3''
to produce files for the LRIS slitmask mill machine and the slitmask 
alignment program. The slit mask allowed to reobserve the same PNs
observed with FORS1. Some slits were assigned to control stars, like
we did with FORS1. The Keck I, LRIS-B observations were made on
the night of 2003 April 2/3. LRIS-B gives an 8 arc min field on a
mosaic of two 2k$\times$4k CCDs (pixel size 
15$\times$15 $\mu$m). The image scale is 0.135 arcsec pix$^{-1}$. The
observing mode was multi-object spectroscopy (MOS) using dichroic filter 
D680 and grism 400/3400. This grism gives a spectral resolution of about 
2 \AA \ at 5000 \AA, for slits 1 arc sec wide. 

The night was clear and of photometric quality, with seeing from 1 to 0.8
arcsec. We completed 11 MOS exposures, each with an exposure time of
1800 seconds, all of them at air masses smaller than 1.6 except the last
one, with an air mass near 2. We also took spectra of the spectrophotometric 
standard star G193-74 (Oke 1990), with exposure times of 150 and 300 seconds.

The LRIS-B spectrograph does not have an atmospheric dispersion compensator.
In this case we took the standard star spectra through the same mask 
used for the PN spectra and at a low air mass. By comparing to a 
science exposure also taken at a similarly low air mass, the effect 
of atmospheric dispersion is minimized. Other science exposures, taken 
mostly at higher air masses, were corrected using the control stars,
as described in the previous section.

The spectra were reduced in a similar way as already described for FORS1. 
The only difference is that we made individual extractions for each
exposure. We did this extra work because LRIS suffers mechanical 
deformations that result in image shifts of several pixels as the telescope 
moves away from the meridian; there was no easy way of registering the
science exposures for combination. This also forced us to spend some time
carefully eliminating cosmic ray events from the individual frames.
After extraction, wavelength calibration, extinction correction, flux
calibration and reduction to rest wavelengths, all spectra for each slit 
were combined, assigning weights proportional to the strength of [O III]
$\lambda$5007. The final fluxes in erg cm$^{-2}$ s$^{-1}$ were obtained by 
integrating the emission lines over wavelength. For comparison with the FORS1
spectra, the higher-resolution LRIS-B spectra were smoothed with a boxcar=7.

\section{Results}

It was reassuring to find that, for each PN, the FORS1 and LRIS-B spectra 
are in good agreement in the wavelength range where they can be compared.
The visible emission lines in that range are H$\beta$, 
[O III] $\lambda\lambda$4959, 5007, and H$\alpha$. 
Figs. 5 to 7 show the comparison of FORS1 vs. LRIS-B spectra. PN1 was not
seen by FORS1; it fell outside the end of the first slit. Some LRIS-B
spectra did not reach H$\alpha$ because of the PN positions in the field
(our fault).
The LRIS-B spectra confirm that many PNs in our sample have very strong
5007/H$\beta$ ratios, and provide clear detection of [Ne III] in most
PNs, as shown in Fig. 8; [O II] $\lambda$3727 is not visible, except 
very probably in PN 5.

Tables 4 and 5 show the measured relative intensities, in the scale
H$\beta$=1, calculated in the following way: first calculate the intensity 
relative to $\lambda$5007, which is the strongest and most reliable
line; then average if two measurements are available; then multiply 
by the ratio 5007/H$\beta$. No correction for reddening has been applied.

We consider first the ratio H$\alpha$/H$\beta$. In most cases it is 
consistent with zero or small reddening. The foreground reddening 
produced by our Galaxy is $E(B-V)=0.03$ (see M\'endez et al. 2001) and 
NGC 4697 shows no evidence of interstellar reddening, except for the dust
disk very close to the center. This leads us to expect a H$\gamma$ 
intensity of about 0.47, marginally detectable in individual PN spectra.
In order to improve the signal we averaged all the PN spectra taken with both
spectrographs from 4200 to 5100 \AA. Fig. 9 shows the resulting composite 
spectrum, where H$\gamma$ is indeed visible with the expected intensity.
Also, at the limit of detection lies a weak [O III] $\lambda$4363. If we
take its intensity as an upper limit for $\lambda$4363, 
its relation to $\lambda$4959 and $\lambda$5007
permits to establish an upper limit for the nebular electron temperature.
We measure a ratio (4959+5007)/4363 of the order of 120; from the Osterbrock
(1989) textbook we obtain $T_{\rm e}=12000$ K. We cannot apply this result to 
any individual nebula, but we can argue that no PN in our sample has a very 
high $T_{\rm e}$, because $\lambda$4363 is not visible in any individual
spectrum, and that for any PN with $T_{\rm e}$ slightly
higher than 12000 K there must be a few with fainter $\lambda$4363 and 
therefore with $T_{\rm e}$ lower than 12000 K.

In Fig. 9 we also measure the average ratio 5007/H$\beta$, which turns out
to be 16. This is a very large value, if we compare with what is observed
throughout the Local Group. The only case we know with a similarly high
average ratio is the M 31 bulge sample of Richer, Stasinska \& McCall
(1999). Let us discuss our result more carefully.
Is it possible that we are being misled by a faint H$\beta$ absorption 
component in the background light, which would induce a fainter H$\beta$
emission? We do not think so, for two reasons: first, the effect would be
even stronger for H$\alpha$; although we admit that one could imagine
high internal reddening in some PNs. 
But second, and most important, many PNs with very strong $\lambda$5007 
are in the outskirts of the galaxy, where there is no significant 
background. In fact Tables 4 and 5 show that the strong $\lambda$5007
is more frequent outside than inside, which is contrary to what we would
expect if such a background contamination were playing a role. 

Note also that the largest source of uncertainty in our 
measurements is the determination of the zero intensity level. Given the 
increase in signal to noise provided by the addition of all the spectra,
to find such a large average ratio of 5007/H$\beta$ confirms that for
some PNs the individual ratio must be larger than the average, because
we see in Tables 4 and 5 some cases of ratios below the average (those are
easier to believe because H$\beta$ is stronger and therefore more reliably
measured). For all these reasons we believe the existence of these high
ratios of 5007/H$\beta$ is well established and we must try to explain it.

Fig. 10 shows the average of all the LRIS-B spectra between 3700 and 4200
\AA. It shows prominently [Ne III] $\lambda\lambda$ 3868, 3967, with the
expected relative intensities, as well as a weak H$\delta$, also with the 
expected intensity. But [O II] $\lambda$3727 is not
clearly present. In fact most of the signal at 3727 is provided by only 
one of our sources, namely PN 5 (see Fig. 8). Note in Table 4 that PN 5
has the lowest intensity of $\lambda$5007, which makes sense; a stronger
$\lambda$3727 and weaker $\lambda$5007 clearly indicate a lower excitation 
class.

\section{Analysis}

Since we cannot detect [O III] $\lambda$4363 in the individual spectra,
and we also lack electron density diagnostics,
we cannot make nebular abundance determinations in the normal way.
However, we have found many examples of very high ratios 5007/H$\beta$, 
and we will derive lower limits for the abundances of those PNs. 
That this is possible can be seen e.g. in Fig. 3 of Dopita, Jacoby \&
Vassiliadis (1992),
which shows the ratio 5007/H$\beta$ as a function of abundance and 
stellar effective temperature. Any PN with 5007/H$\beta$ $>$ 18 must 
have a metallicity near solar. We can illustrate this conclusion in
more detail using the CLOUDY code (Ferland et al. 1998). The use of
CLOUDY for PN abundance determinations has been very well described
by Jacoby \& Ciardullo (1999). Here we will simply outline the essential 
steps in our procedure. In Fig. 11 we show our own version of Dopita et 
al.'s Fig. 3. From our Fig. 11, built using CLOUDY, it is obvious that
given low $\lambda$5007 intensities we cannot say much about the O
abundance. We will start our analysis with the ``very strong 
$\lambda$5007'' PNs in our sample, namely PNs 1, 3, 7, 9, 10, 11, 13, 14.

In our spectra we can set upper limits on the intensities of several
important lines relative to H$\beta$: [O II] $\lambda$3727 $<$ 0.4; 
[O III]$\lambda$4363 $<$ 0.3; He II $\lambda$4686 $<$ 0.2. 

These upper limits for faint lines are based on the detectability of 
stronger lines like H$\gamma$ and [Ne III] $\lambda$3967. 
For example, H$\gamma$ $\lambda$4340 is barely seen in most PN spectra, 
at an average intensity of 0.47, while $\lambda$4363 is not seen in any of 
them. The average intensity of $\lambda$4363 on the composite PN 
spectrum (Fig. 9) is at most 0.18. Therefore we estimate that the 
upper limit for $\lambda$4363 cannot be higher than 0.3. In a similar way 
we estimate upper limits for $\lambda\lambda$4686, 3727 (using in this 
last case the detectability of [Ne III] $\lambda$3967 in most PN spectra 
at 0.6, always relative to H$\beta$).

Our purpose is to obtain a lower limit to the O and Ne abundances 
by adjusting several nebular and stellar parameters (nebular density 
and radius, stellar luminosity
and temperature) until we obtain a CLOUDY run whose line intensities 
reproduce the observed O and Ne lines and do not violate any of the 
intensity limits we have determined for the weaker lines.

After a few dozen CLOUDY runs we converged to the following solution: a
small nebular radius, 0.017 pc, and high electron density, 4$\times 10^4$
cm$^{-3}$, both needed to keep $\lambda$3727 at low intensity; a central star 
luminosity of at least 7000 $L_\odot$ (it will be higher if the nebula
leaks some H-ionizing photons), required to reproduce the nebular H$\beta$ 
luminosity, $\sim 10^{35}$ erg s$^{-1}$, which we know from the
distance to NGC 4697 and the $\lambda$5007 flux (M\'endez et al. 2001)
plus the 5007/H$\beta$ ratio measured in our spectra; and a stellar
temperature not higher than 10$^5$ K, in order to keep He II $\lambda$4686
at low intensity. With these parameters, and assuming that these PNs have
typical dust grain contents, CLOUDY confirms that in order to produce 
$\lambda$5007 and $\lambda$3868 intensities of 19 and 1.9, respectively,
we need O and Ne abundances of 8.60 and 7.88, in the usual logarithmic scale 
where H=12. Lowering the central star temperature or the nebular density (if
we allow for a stronger $\lambda$3727) would require to increase the 
abundances. A hotter star, apart from producing a too strong
$\lambda$4686, would also require higher abundances. Therefore we are 
confident that the O and Ne abundances we derived are lower limits. We
show this in Table 6, where we list a selection of CLOUDY runs illustrating
the effects of departing from the chosen values of central star surface
temperature, nebular density and nebular radius (the last two parameters
have no significant influence on the abundances).

The objects PN1 and PN3 have a 5007/H$\beta$ ratio of 22; in this case
the lower limit to the O abundance is 8.75.

Now we consider O abundances in PNs 2, 4, 5, 6, 8, 12. A weak $\lambda$5007 
is not inconsistent with metallicities higher than solar (see e.g. our 
Fig. 11, or Fig. 3 in Dopita, Jacoby \& Vassiliadis 1992). Therefore we 
cannot set any upper limit. On the other hand,
we know that none of these PNs has a very high 
$T_{\rm e}$, because we could not detect [O III] $\lambda$4363 in 
individual spectra. This rules out {\it very\/} low O abundances, 
but we cannot exclude values around 8.2. So we set that number as
a lower limit for the six PNs with weaker $\lambda$5007. 
These six PNs will not provide stronger constraints 
until we can detect, in their individual spectra, 
the diagnostic lines required for accurate abundance determinations. 

\section{Interpretation of the O and Ne abundances from PNs}

\subsection{Comparison with the absorption-line data}

Fig. 12 shows the abundance limits derived from the PNs, together with
the absorption-line information presented previously in Fig. 3.
The absorption-line abundance gradient is derived in total metallicity, 
which is dominated by the ``alpha'' elements. 
More specifically, the contribution of iron-peak elements
to the total metallicity is about 10-15\%, while alpha-elements,
including oxygen, contribute more than 70\%. In our modelling, 
$[\ZH]$ is a well-defined quantity and it means real total 
metallicity, i.e. all elements heavier than helium. Please refer to 
Thomas, Maraston, \& Bender (2003) for more details.

Therefore our nebular O/H should be 
compared to the $[\ZH]$ in Fig. 3. In addition, we need to specify what is 
the solar abundance of oxygen. From Lodders (2003) we adopt a solar O
abundance 8.7 (logarithmic, in the scale where H=12). Then for example our
O abundance 8.6 becomes $[\ZH] = -0.1$.

In Fig. 12 we find that at least PNs 1, 3, 7, 9, 10, 11, 13, 14 have
higher metallicities than expected from the integrated absorption-line 
spectra. This is direct evidence of a 
metal-rich population, both within and beyond one effective radius 
from the center of NGC 4697. Then we have the six
PNs with very low lower limits. If their real abundances were all close to 
the lower limits, then the distribution would be markedly bimodal. We find
this implausible, and we rather expect most of their abundances to lie much
closer to the upper PN group in Fig. 12 (which are, please note, also lower 
limits), perhaps producing an average 
abundance not very different from what the absorption-line data indicate for
$\log r = 1.5$. This would require the rather steep metallicity gradient near 
the core of NGC 4697 to become much less pronounced beyond one effective 
radius.

However, before declaring the PNs to be representative of the stellar 
population, we need to discuss if the rather high average PN metallicities 
can be attributed to a selection effect. 

\subsection{On selection effects and metallicity distributions}

Concerning PNs there is an obvious selection effect: we have restricted
our sample to objects populating the high-luminosity end of the PN 
$\lambda$5007 luminosity function. Are their O abundances higher than 
those of fainter PNs? This has been discussed e.g. by Richer (1993), 
Stasinska, Richer \& McCall (1998), Jacoby \& Ciardullo (1999), and 
Magrini et al. (2004). 

In theory we expect high O abundance to produce a stronger $\lambda$5007 
emission line. On the other hand, more metallicity could imply more AGB 
mass loss (e.g. Greggio \& Renzini 1990) and a lower central star 
luminosity, which would then cancel the O abundance effect mentioned 
earlier. Since our theoretical 
knowledge of the influence of metallicity on AGB and post-AGB evolution
is far from ideal (mostly because we lack a good theory of AGB mass loss
processes) it looks safer to rely on the available empirical information:
a direct comparison of the average PN abundances in  
progressively weaker bins along the luminosity function. This can be done
only for nearby galaxies, like the Magellanic Clouds or M 31. 

The main argument favoring an almost solar metallicity population 
in the outskirts of NGC 4697 is the fact that we found so many PNs 
(80\% of those beyond one effective radius) with very strong 5007/H$\beta$. 
As mentioned before, this happens in only one other place throughout our
Local Group: the bulge of M 31 (Richer, Stasinska \& McCall 1999). 
Combining the information given in their Tables 5 and 9 (oxygen abundances 
and line intensities, respectively, in the bulge of M 31) with the 
apparent $\lambda$5007 magnitudes from the discovery paper (Ciardullo et 
al. 1989) we find for the 16 brightest M 31 bulge PNs that 9 of them, or 
56\%, have 5007/H$\beta$ $>$ 18, and we find that the average oxygen 
abundance of these 16 PNs is at least 8.70 (there are 4 lower limits). 

We can test how important it is to be restricted to the most luminous 
PNs; taking the total sample of 28 PNs we get an average O abundance of 
at least 8.64, with 14 lower limits. There does not seem to be a very 
significant difference. But what happens if we consider even fainter PNs?
We turn to Jacoby \& Ciardullo's (1999) analysis of the M 31 PNs. First,
their Fig. 6 shows that across one order of magnitude in oxygen 
abundance, from 8.1 to 9.1, the upper envelope is flat; i.e. a 
metal-poor PN can be as bright as a metal-rich PN. This is an essential 
point: if there is a population with LMC or SMC metallicity in NGC 4697, 
it will produce bright PNs and we will find them (very probably we did, 
in fact, but we cannot be sure because a low 5007/H$\beta$ admits many 
possible abundance interpretations).

On the other hand, Jacoby \& Ciardullo remark that their fainter M 31 
PNs have lower 
average metallicity. However, they do not conclude that the brighter PNs 
must overestimate the metallicity; in their Section 4.4.1 they comment
that the tendency for fainter PNs to be drawn from a lower-metallicity 
population may be an artifact of a selection effect or an analysis bias, 
because there is no physical rationale why metal-rich PNs should not be 
faint as well as bright. 

This interpretation is supported by the fact that the average "brightest 
PN" abundance (from the data in Richer, Stasinska \& McCall 1999) agrees 
with another set of data: absorption-line indices yield super-solar 
metallicities for the integrated stellar population of the M31 bulge 
(Trager et al. 2000a, 2000b; Thomas et al. 2005). If we expected the 
brightest PNs to overestimate the M31 bulge metallicity, then we would 
expect the M 31 integrated light to indicate a lower average metallicity. 
Since that does not happen, Jacoby and Ciardullo 
feel that they need to explain why their sample has so few metal-rich 
objects, coming from a metal-rich population. The brighter Richer et al.
sample does not have that problem. 

Coming back to NGC 4697,
we still have to discuss another possibility, involving now a very low 
metallicity. Consider the study by Richer (1993) of the LMC and SMC PNs. 
In his Fig. 2 we find that the more metal-rich objects are brighter; not by 
much initially, but the difference becomes important when we reach oxygen 
abundances below 8.0. Then, if there were many very-metal-poor PNs
in the outskirts of NGC 4697, they would be fainter and not detectable 
in our flux-limited sample. But then we would be talking about a peak 
in $[\ZH]$ below $-1$, and therefore about an extremely bimodal metallicity 
distribution. A gaussian metallicity distribution (or even a moderately 
bimodal one, as
sometimes observed in globular cluster systems, e.g. Zepf \& Ashman 1993;
Gebhardt \& Kissler-Patig 1999) extending that far down in metallicity would 
require a lot of LMC- and SMC-like PNs, which we would detect, as argued 
above, so that the frequency of very strong 5007/H$\beta$ values would 
not be as high as we observe. 

Since the observational evidence is not overwhelming, we do not rule out
that our bright PNs could have $[\ZH]$ higher by about 0.2 than the
stellar population they represent. This would bring the average 
metallicity, as derived from the PNs, 
in even better agreement with the long-slit results inside 
one effective radius. We emphasize, however, that right now there 
is no compelling theoretical reason nor any evidence in M 31 (the only 
testing ground we have) indicating that such a systematic difference
between PNs and stars must necessarily exist.

But even assuming such a difference, our basic conclusion remains valid: 
a similar metallicity both inside and outside 
of one effective radius requires the abundance gradient to become less 
steep outside. Alternatively, the conceivable existence of a very
metal-poor population with $[\ZH]$ below $-1$, which we cannot
exclude, would require an extremely bimodal metallicity distribution.
Let us remark, in closing this discussion, that the metal-rich 
population represented by our PNs is quite real, while the 
very metal-poor population must remain hypothetical until confirmed
by future observations. The sad fact is that based only on our PN lower 
limits we cannot estimate how large is the dispersion in abundances.
Abundance determinations for fainter PNs in NGC 4697 are probably beyond 
the reach of currently available instrumentation and may require a 
30m telescope.

\subsection{Other galaxies}

The PNs in
NGC 5128 seem to have significantly lower metallicities, well below
solar (Walsh et al. 1999). This is consistent with the metallicity
distribution in the halo of NGC 5128 derived by Harris \& Harris
(2002), which indeed does not have a prominent tail of stars with
super-solar metallicities. On the other hand the information about
PN metallicities in NGC 5128 is still very limited and uncertain;
it seems prudent to postpone any further discussion until more and
better PN abundances are measured in NGC 5128.

Consider now PNs in the bulge of the Milky Way. Ratag et al. (1992, 1997) 
derive an oxygen abundance distribution for PNs in our Galactic bulge 
which, allowing for the lower oxygen solar abundance we favor today, 
seems consistent with the metallicity distribution of field stars
(Zoccali et al. 2002), which peaks near solar metallicity.

Unfortunately it is difficult to build a flux-limited PN sample in the
Milky Way bulge, because of the strong reddening and the depth of the 
bulge. We note that in the Milky Way bulge there are few 
examples of very high ratios 5007/H$\beta$ (see Ratag 1991), but we 
will refrain from interpreting this fact.


NGC 4697 does not belong among the most luminous and redder ellipticals.
It now becomes interesting to investigate if other elliptical galaxies, 
particularly the redder ones, show a high frequency of ``very strong 
$\lambda$5007'' PNs. In fact, it would be prudent also to verify what 
happens on the other side of NGC 4697. This kind of work, requiring
only the detection of H$\beta$, is not so demanding, in terms of 
big telescope time, as an accurate abundance determination.

\section{Summary and perspectives}

This paper is a first attempt to use PNs for the determination of 
individual stellar abundances in elliptical galaxies. For the first time
we have taken deep spectra of a relatively large sample of 14 PNs at a 
variety of angular distances from the center of a typical elliptical
galaxy, NGC 4697, located at about 11 Mpc from us. The most important
result is the high frequency of PN spectra with very strong
5007/H$\beta$ ratios $\sim$ 20.
We argue that such ratios require O abundances near solar, 
or may be even above solar. The Ne abundances are similarly high. 
The high O and Ne abundances are direct evidence of the existence of a 
metal-rich stellar population, confirming in {\it individual\/} stars
the high metallicities derived in many cores of elliptical galaxies 
from absorption-line indices measured on integrated-light, long-slit 
spectra (e.g. Trager et al. 2000a, 2000b; Thomas et al. 2005).
This result also validates the derivation of metallicities in early-type
galaxies from the direct comparison with Galactic globular clusters
(Maraston et al. 2003).

Perhaps more surprising is the existence of a metal-rich population 
extending as far as more than two effective radii from the center of 
NGC 4697. This forces us to conclude that the abundance gradient, 
which is rather steep within one effective radius, must become much less 
pronounced outside. The only alternative to this interpretation would be 
the presence of a very metal-poor PN population with $[\ZH]$ below $-1$,
which we cannot rule out but would require an extremely bimodal
metallicity distribution with almost no intermediate metallicities.

It would now be interesting to verify 
what is the situation on the other side of NGC 4697, as well as in other 
elliptical galaxies.

High metallicities in elliptical galaxies imply not very high electron 
temperatures, and therefore very weak [O III] $\lambda$4363
intensities, making abundance determinations in the normal way very costly
in terms of 10m-class telescope time. However, we have found that the
ratio 5007/H$\beta$ can provide important constraints on the abundances
and is worth exploring systematically in the many elliptical galaxies
where PNs have already been discovered. 

We thank the anonymous referee for several useful comments.

\newpage

\appendix

\section{APPENDIX: finding charts}

In Figs. 13 and 14 we provide finding charts for the 14 PNs studied in this 
work, taken from the discovery on-band combined image in M\'endez et al. 
(2001).
Each chart is approximately 25 $\times$ 25 arc seconds, with only one
exception: the last one, showing PN13 and PN 14, which is 50 $\times$ 50
arc seconds. In all cases the orientation is 
the same as in M\'endez et al. (2001). The North direction is obtained by 
rotating 24 degrees counterclockwise from the vertical direction.

\newpage

\figcaption[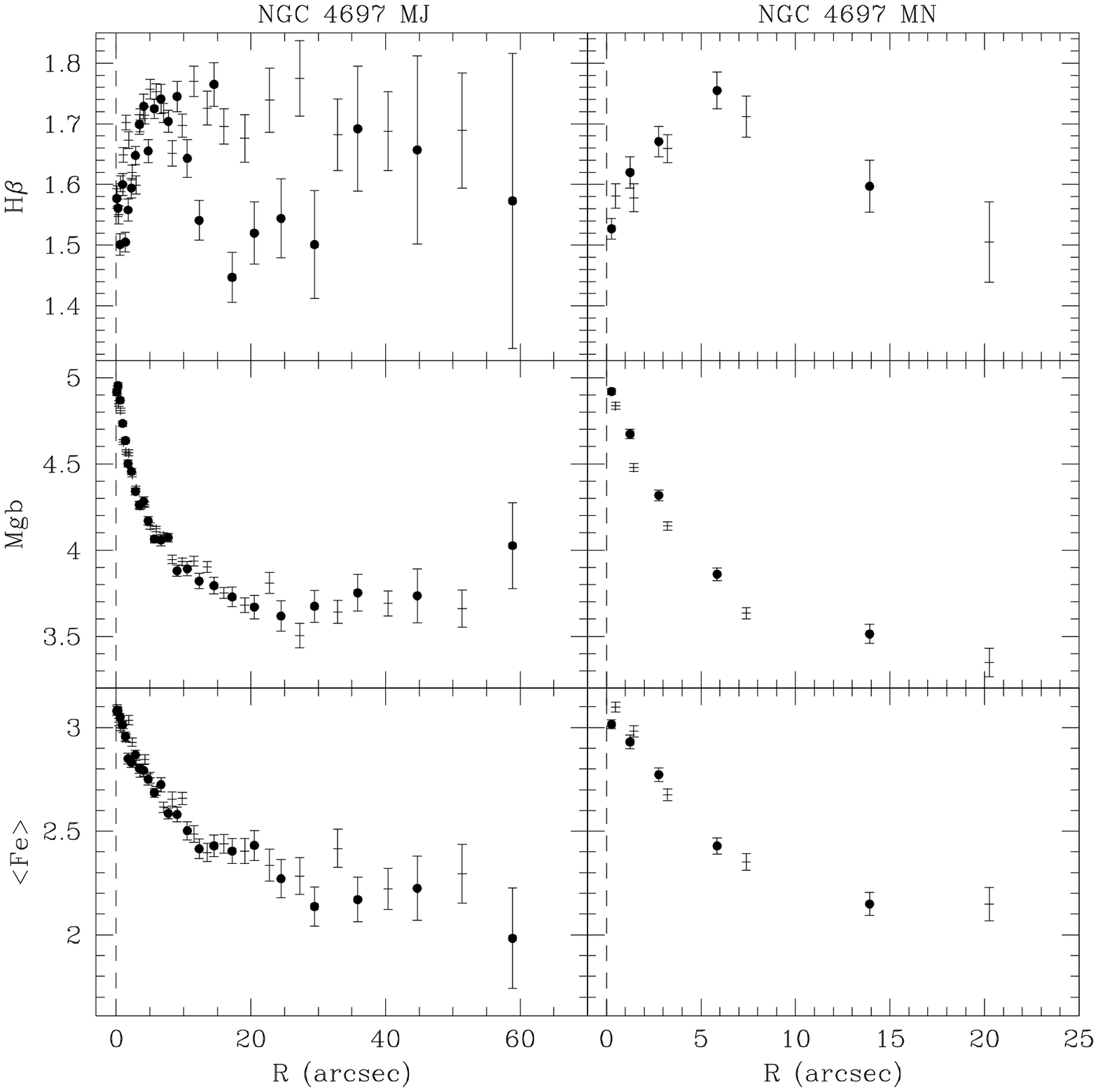]{The strengths of the line indices \Mgb, \Fe\
and \Hb, as a function of the distance in arcsec from the center of the 
galaxy, along the major (left) and the minor (right) axes of NGC 4697. 
Filled circles are positive radii; the positive sense is defined  
towards NE along the major axis, and towards NW along the minor axis.
\label{fig1}}

\figcaption[f2.ps]{Lick absorption line indices \MgFe\ vs.\ \Hb\
(left panel) and \Mgb\ vs.\ $\Fe=0.5 ({\rm Fe5270}+{\rm Fe5335})$
(right panel).  Overplotted are the TMB03 stellar population models
for various ages, metallicities, and \aFe\ ratios as indicated by the
labels in the diagrams. In the left panel, \aFe\ is solar, while in
the right panel the age is fixed to $12\;$Gyr. Open diamonds and 
filled squares are along the major and minor axes, respectively.
\label{fig2}}

\figcaption[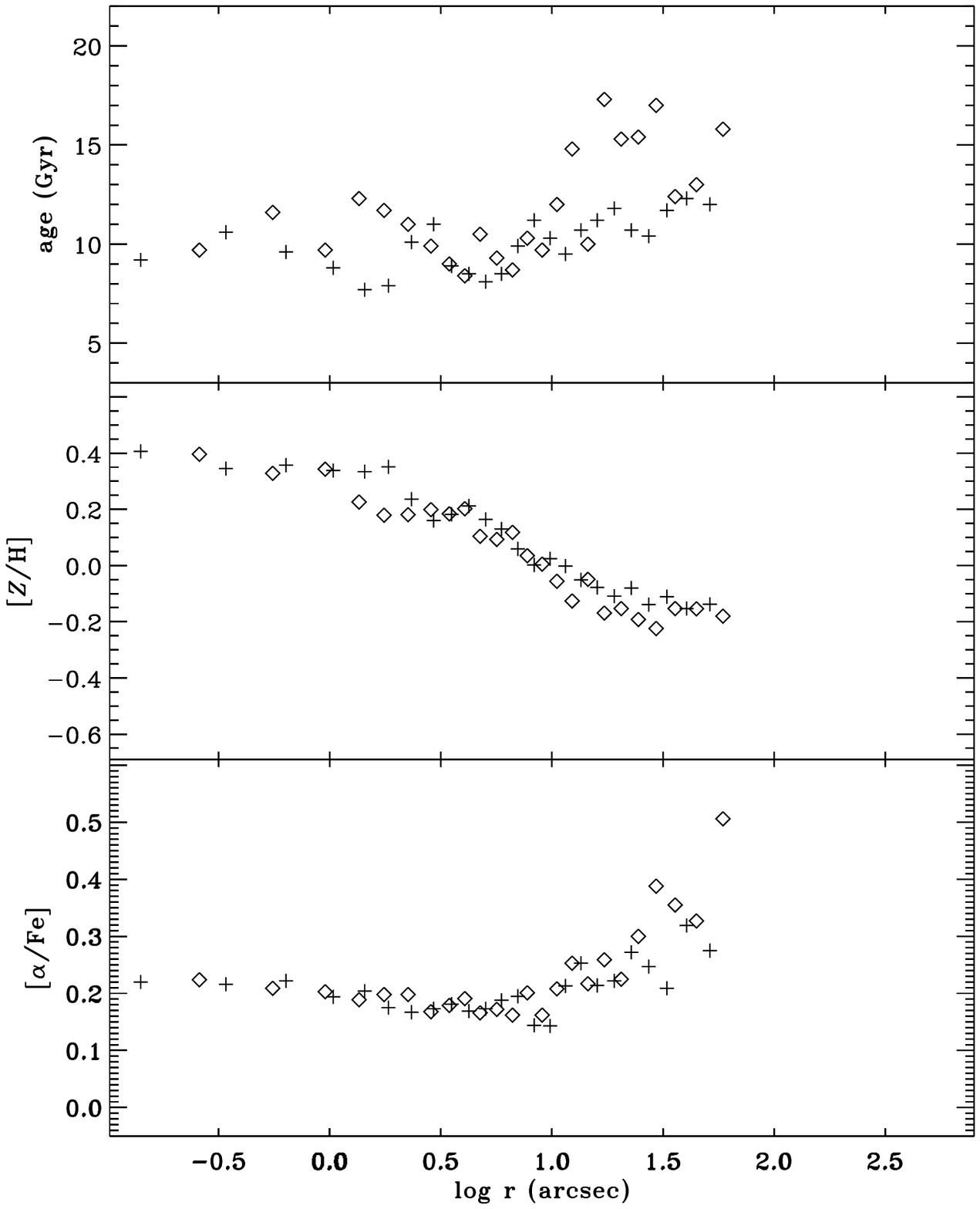]{Stellar population parameters derived from Fig. 2 as
functions of galactic radius. The effective radius is at $\sim
95\;$arcsec. Diamonds and plus signs are positive and negative
radial directions, respectively, along the major axis.
\label{fig3}}

\figcaption[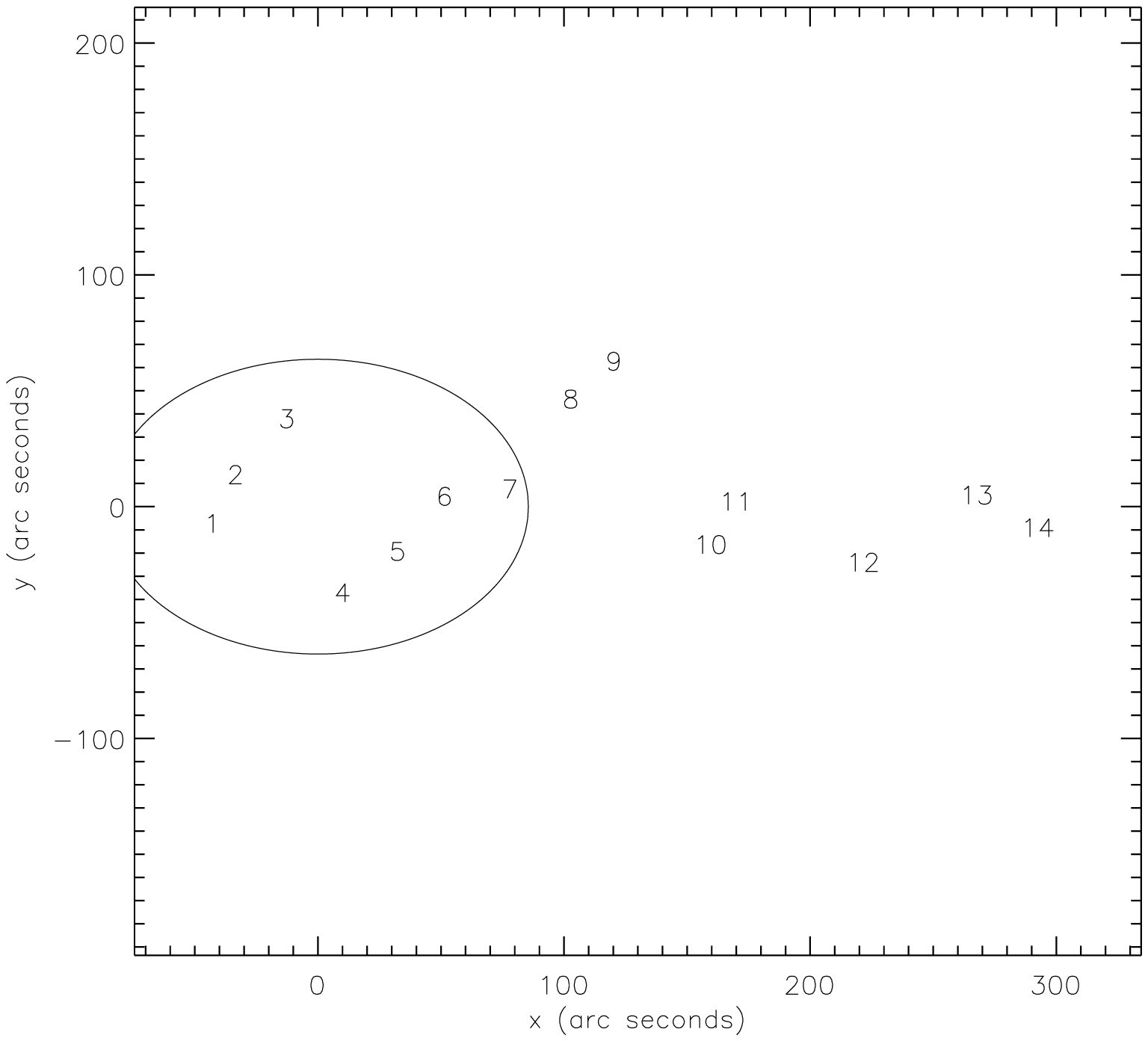]{Positions of the 14 bright PNs relative to the core 
of NGC 4697. The area and orientation of this figure are as in Fig. 7 of
M\'endez et al. (2001). The ellipse is a schematic representation of
NGC 4697. The origin of coordinates is located at the light center of
the galaxy. The effective radius of NGC 4697 is 95 arc seconds. 
\label{fig4}}

\figcaption[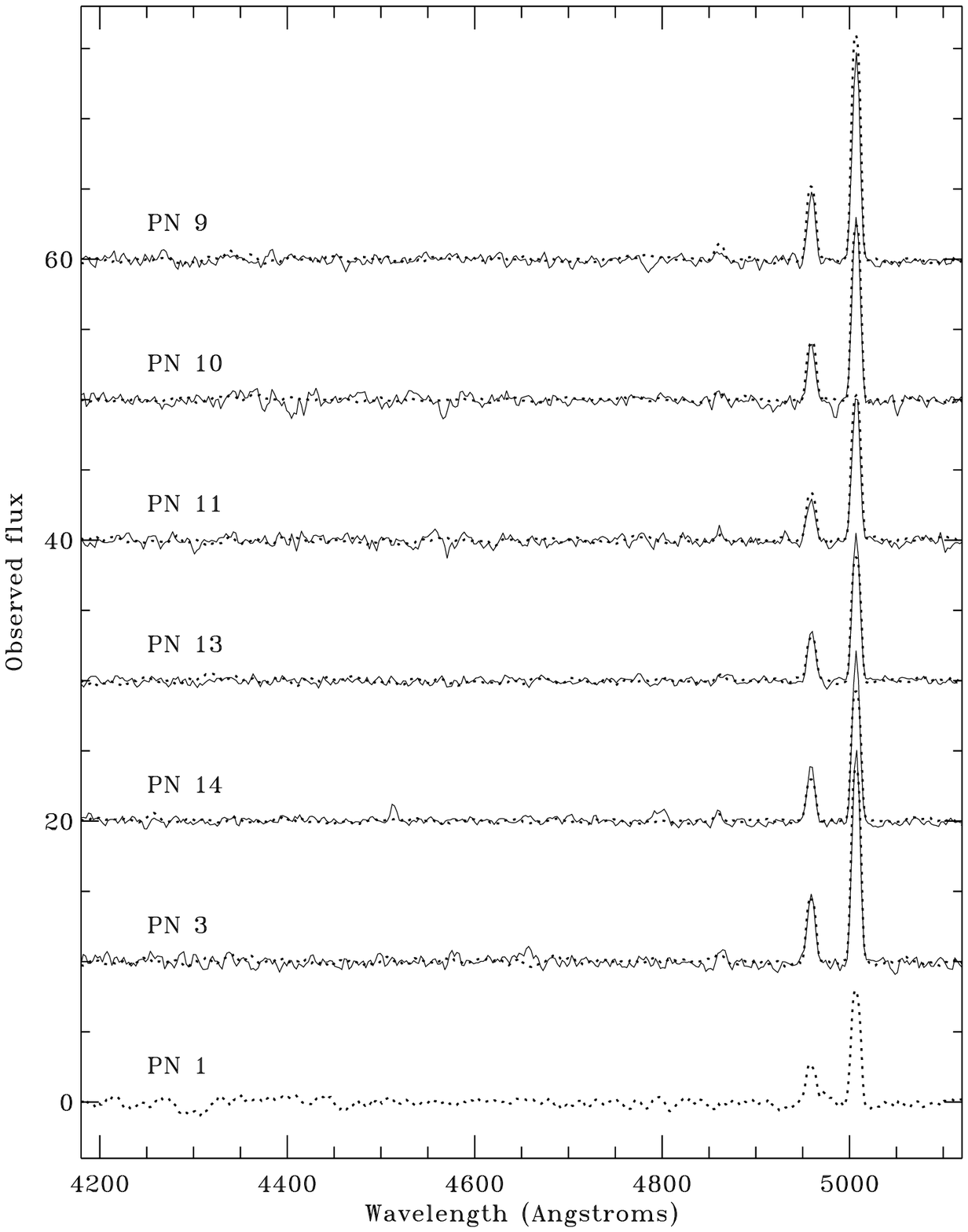]{Comparison of individual PN spectra taken with FORS1
(full lines) and LRIS-B (dotted lines). The PNs are designated with 
the numbers shown in Fig. 4. The fluxes are expressed in units of
10$^{-18}$ erg cm$^{-2}$ s$^{-1}$ \AA$^{-1}$. The levels of zero
intensity are separated by 10 flux units for easier comparison.
The higher-resolution LRIS-B spectra have been smoothed with a boxcar=7.
PN 1 could not be observed with FORS1. 
\label{fig5}}

\figcaption[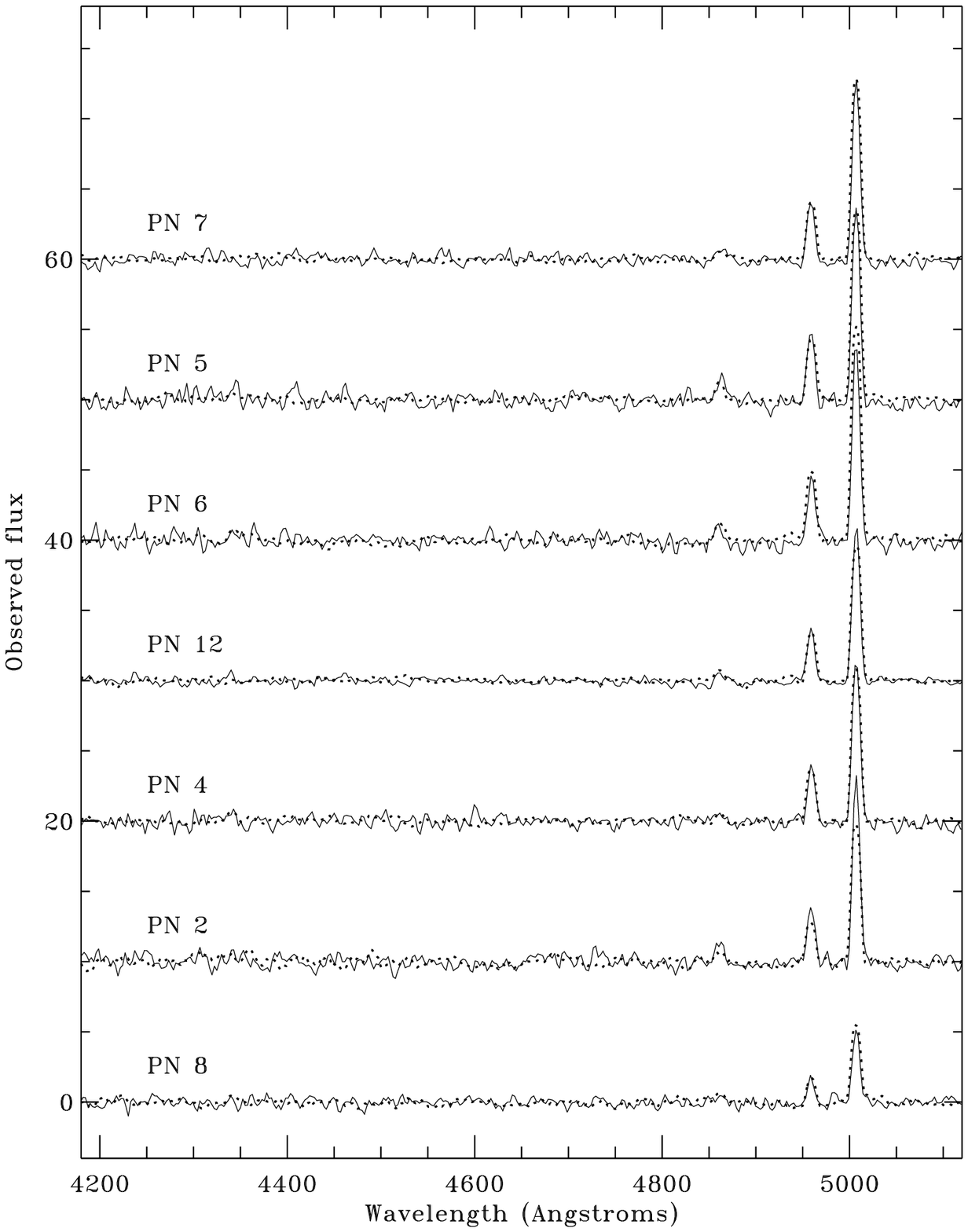]{Comparison of individual PN spectra taken with FORS1
(full lines) and LRIS-B (dotted lines). The PNs are designated with 
the numbers shown in Fig. 4. The fluxes are expressed in units of
10$^{-18}$ erg cm$^{-2}$ s$^{-1}$ \AA$^{-1}$. The levels of zero
intensity are separated by 10 flux units for easier comparison.
The higher-resolution LRIS-B spectra have been smoothed with a boxcar=7.
\label{fig6}}

\figcaption[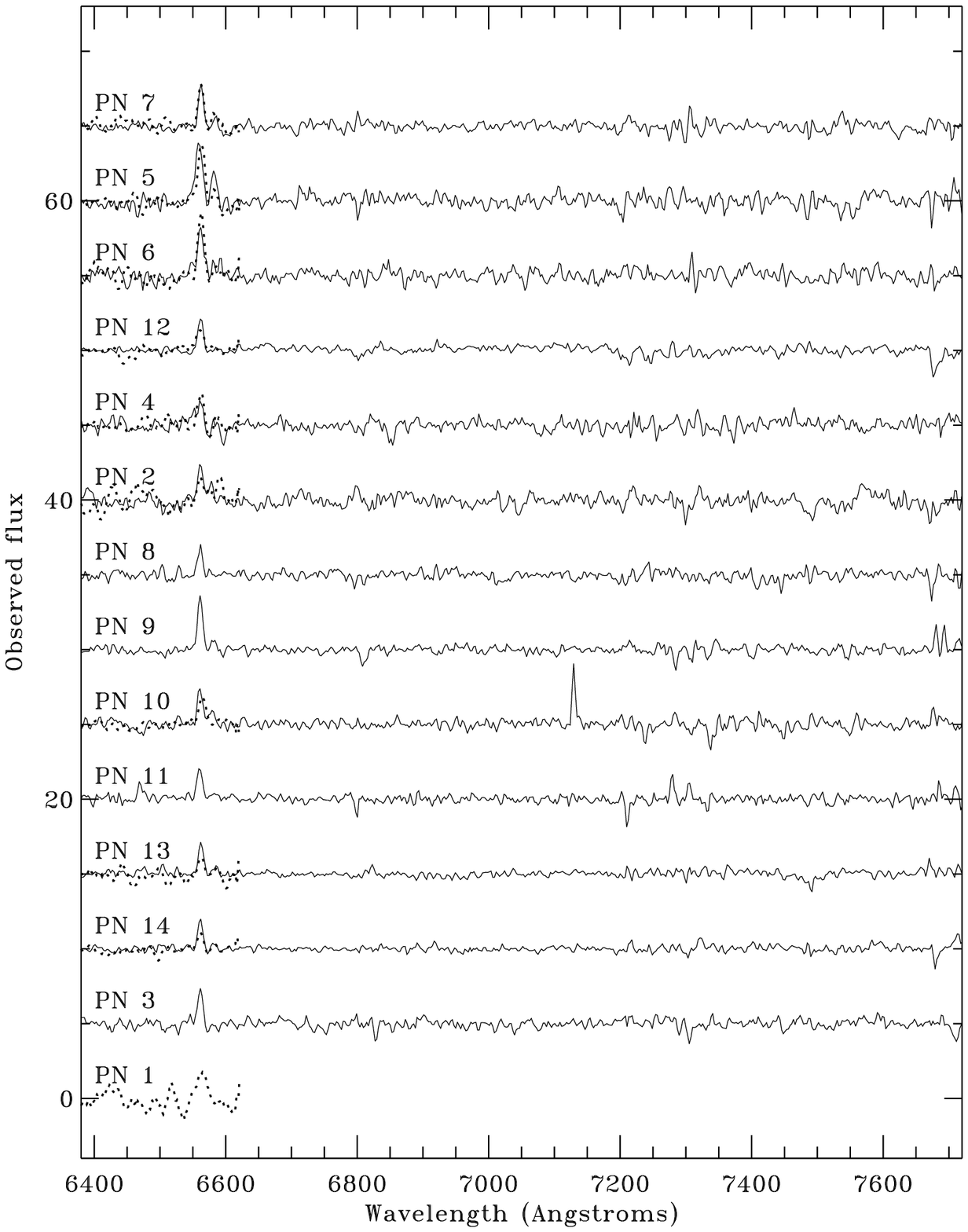]{Comparison of individual PN spectra taken with FORS1
(full lines) and LRIS-B (dotted lines). The PNs are designated with 
the numbers shown in Fig. 4. The fluxes are expressed in units of
10$^{-18}$ erg cm$^{-2}$ s$^{-1}$ \AA$^{-1}$. The levels of zero
intensity are separated by 5 flux units for easier comparison.
The higher-resolution LRIS-B spectra have been smoothed with a boxcar=7.
The feature longward of H$\alpha$ 6563 is not [N II] $\lambda$6584; it is 
due to residual sky emission. The sky emissions are blueshifted because
the PN emission lines have been shifted to their rest values.
\label{fig7}}

\figcaption[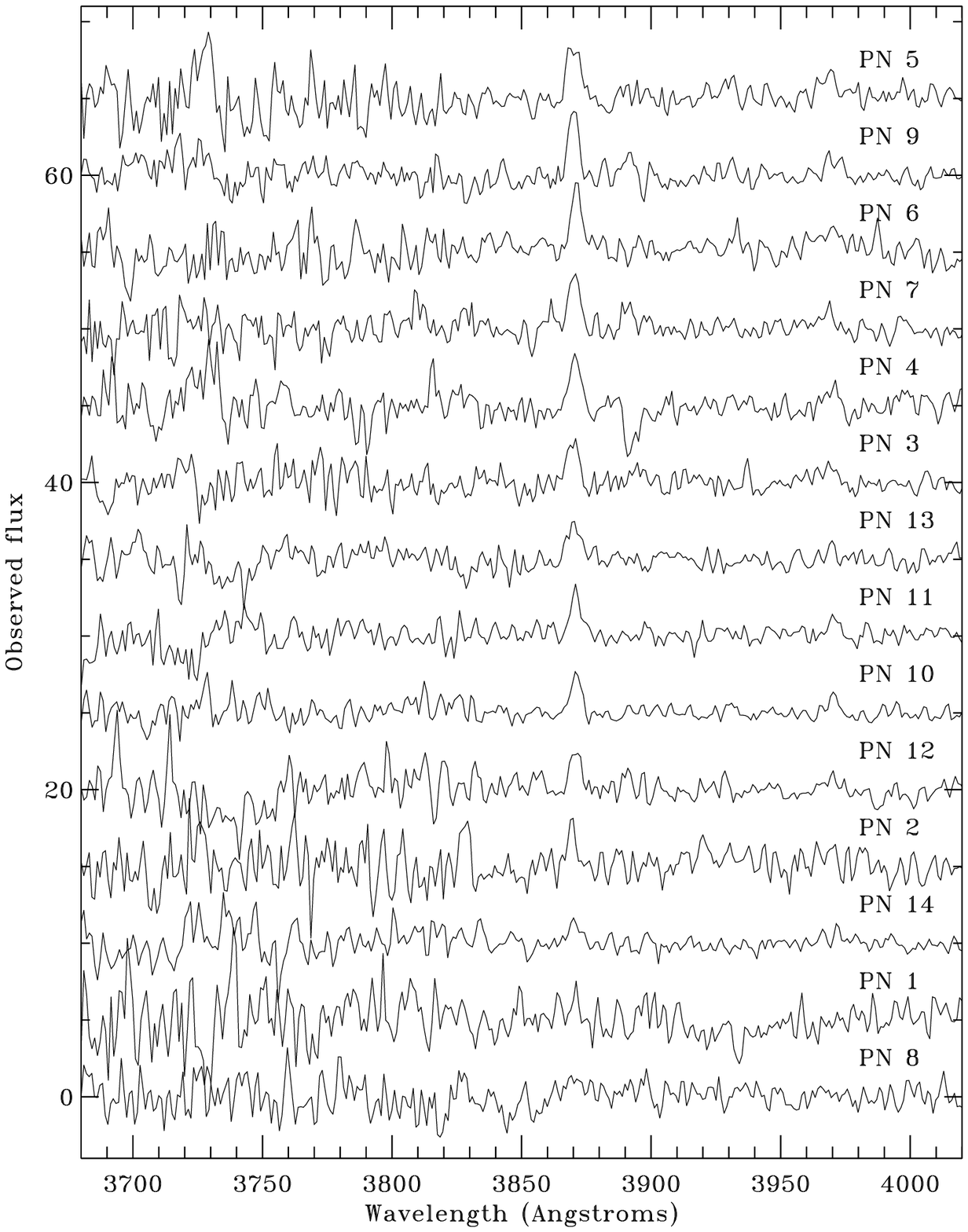]{Individual LRIS-B PN spectra from 3700 to 4000 \AA.
The fluxes are expressed in units of 
10$^{-18}$ erg cm$^{-2}$ s$^{-1}$ \AA$^{-1}$. The levels of zero
intensity are separated by 5 flux units for easier comparison.
The visible features are [Ne III] $\lambda$3868 and (marginally) 
[Ne III] $\lambda$3967. PN 5 may show some [O II] $\lambda$3727.
\label{fig8}}

\figcaption[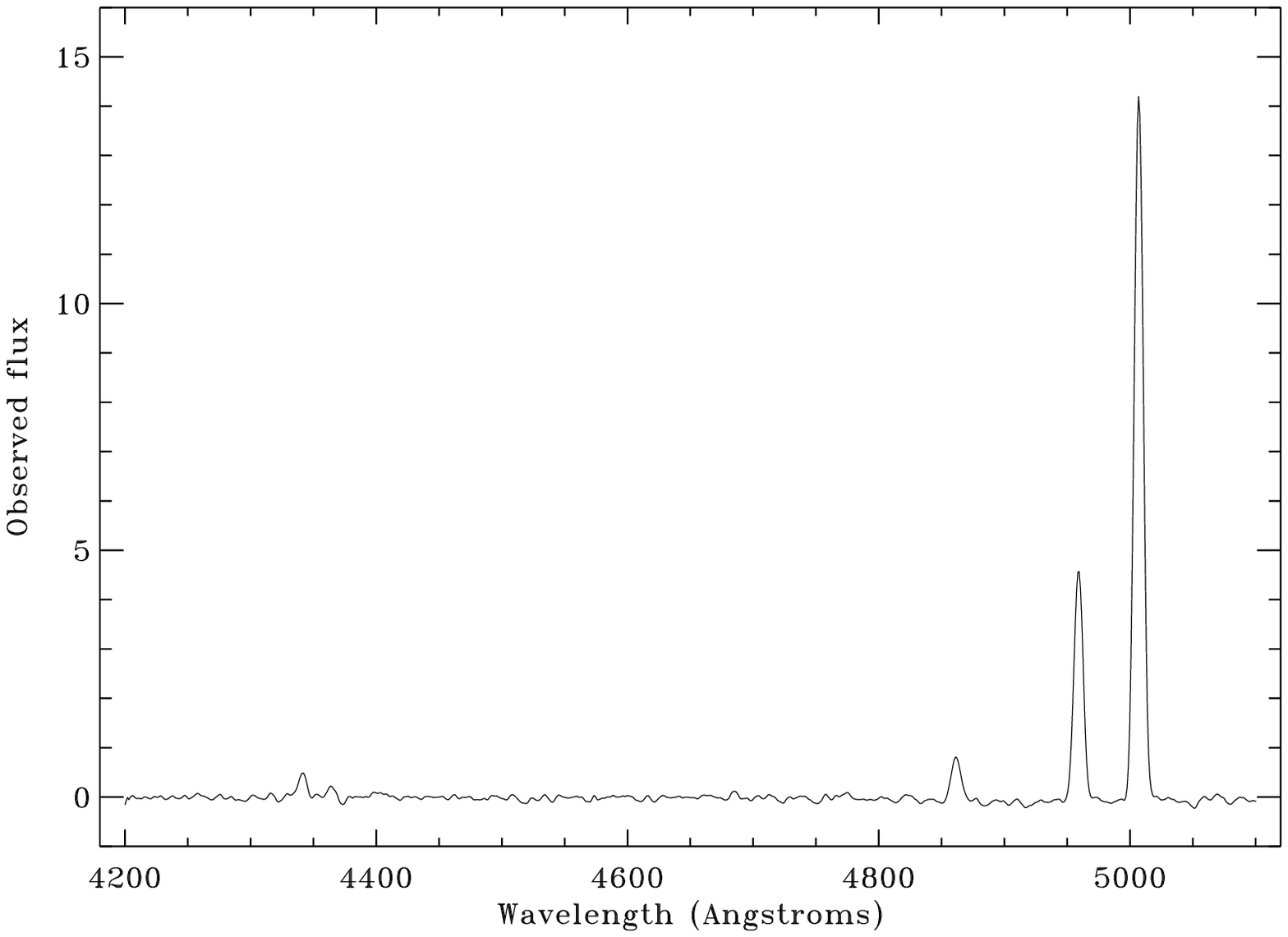]{Average of all spectra of the 14 PNs in NGC 4697.
Visible features are H$\gamma$ $\lambda$4340, [O III] $\lambda$4363,
possibly a very weak He II $\lambda$4686, H$\beta$ $\lambda$4861, [O III]
$\lambda\lambda$4959, 5007.
\label{fig9}}

\figcaption[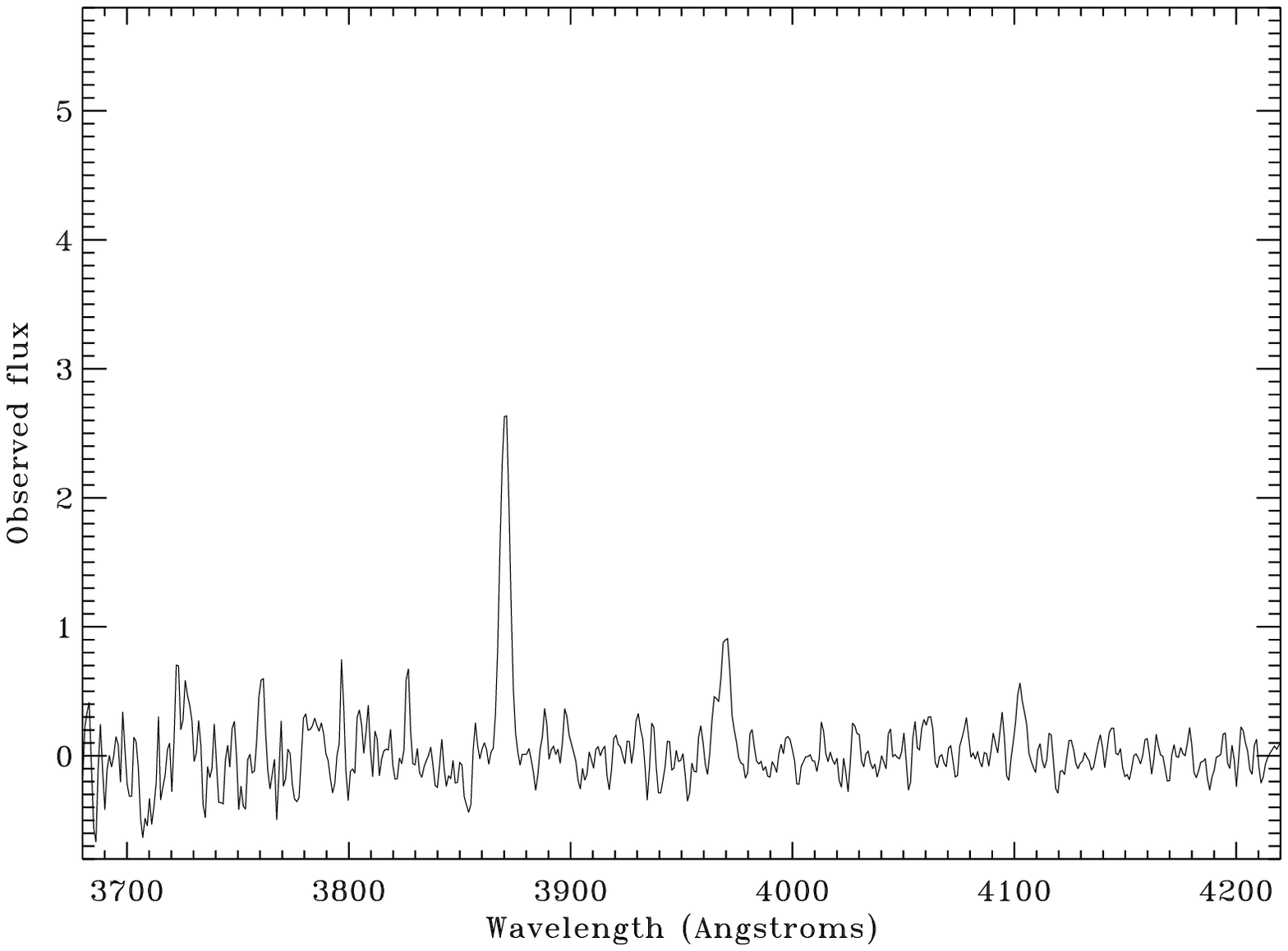]{Average of all LRIS-B spectra of the 14 PNs in NGC 4697.
Visible features are [Ne III] $\lambda\lambda$3868, 3967,
H$\delta$ $\lambda$4101.
\label{fig10}}

\figcaption[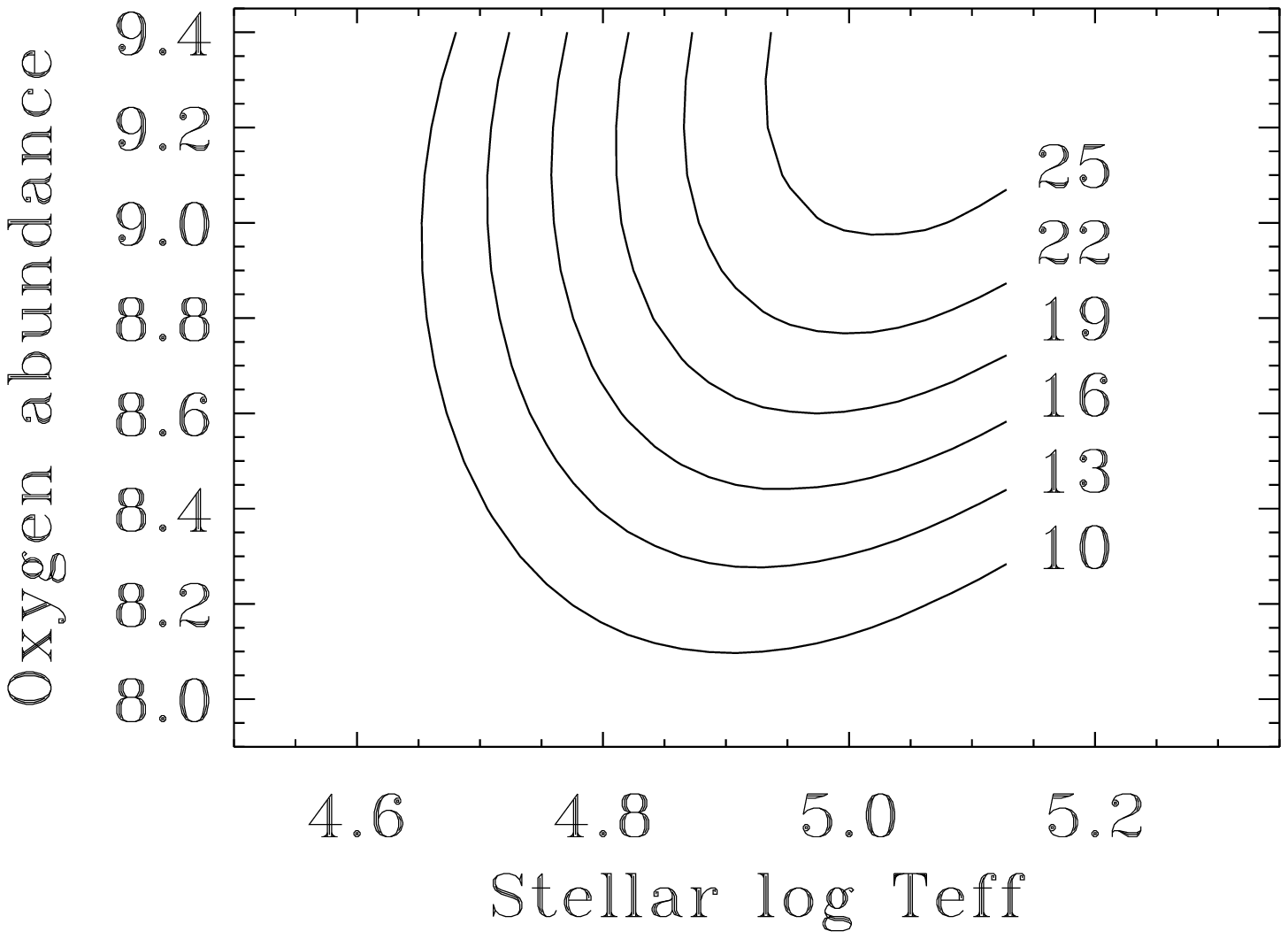]{This is our imitation of Dopita et al.'s Fig. 3 
(see text). Our grid was calculated using CLOUDY. 
The contours give the 5007/$H\beta$ ratio as a function of stellar 
effective temperature (we assume blackbody energy distributions) and
oxygen abundance (logarithmic, in the scale where H=12).
\label{fig11}}

\figcaption[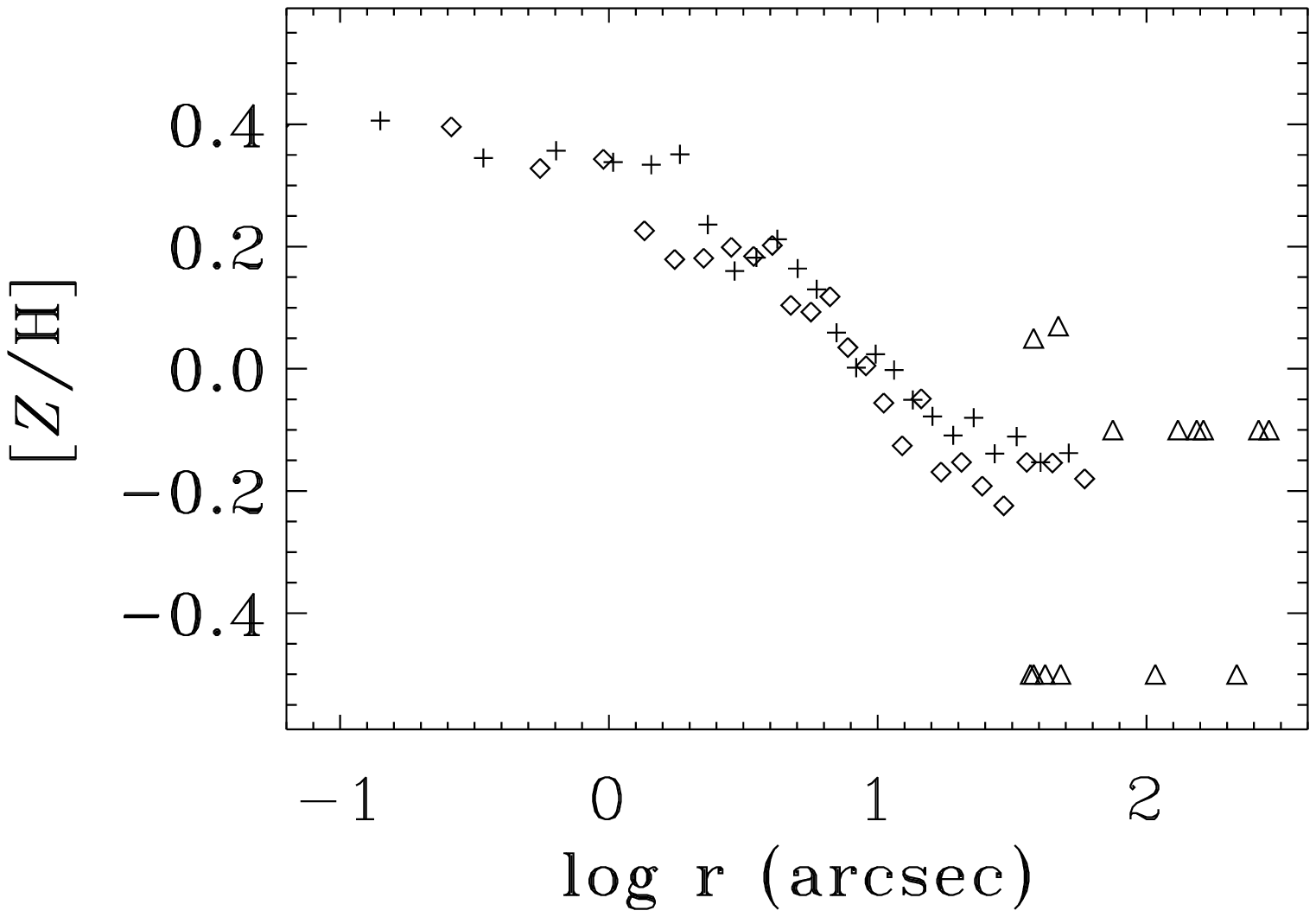]{The combined information about metallicity from 
absorption-line data (diamonds and plus signs as in Fig. 3) 
and from the PNs (triangles pointing upward indicate lower 
limits to the abundances) plotted as a function
of projected distance to the center of NGC 4697. One effective radius
is at $\log r = 1.98$.
\label{fig12}}

\figcaption[f13.ps]{Finding charts for the PNs studied in this work.
\label{fig13}}

\figcaption[f14.ps]{Finding charts for the PNs studied in this work.
\label{fig14}}

\newpage

\begin{figure}
\figurenum{1}
\epsscale{1.0}
\plotone{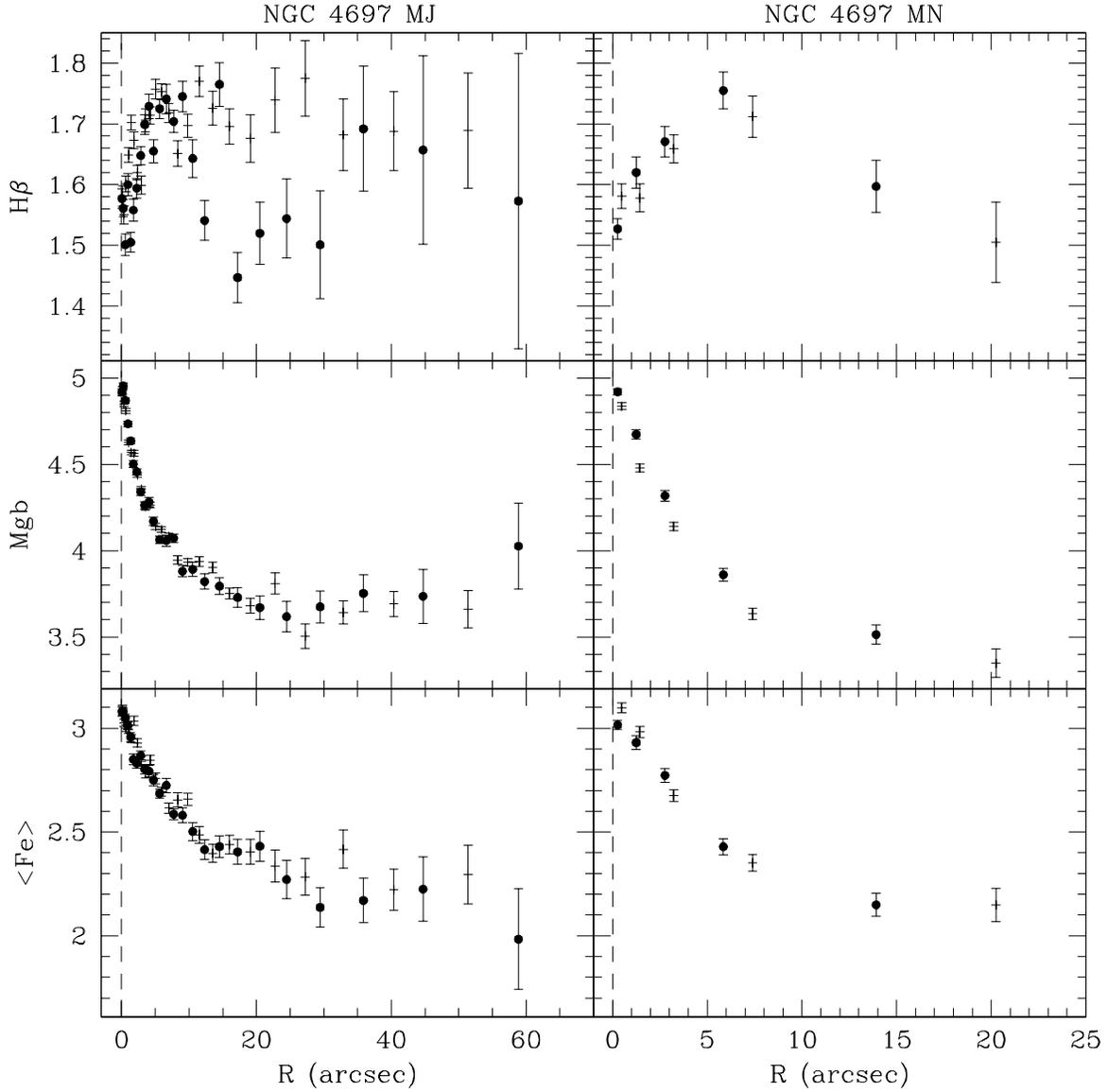}
\caption{The strengths of the line indices \Mgb, \Fe\
and \Hb, as a function of the distance in arcsec from the center of the 
galaxy, along the major (left) and the minor (right) axes of NGC 4697. 
Filled circles are positive radii; the positive sense is defined  
towards NE along the major axis, and towards NW along the minor axis.}
\end{figure}

\begin{figure}
\figurenum{2}
\epsscale{1.0}
\plottwo{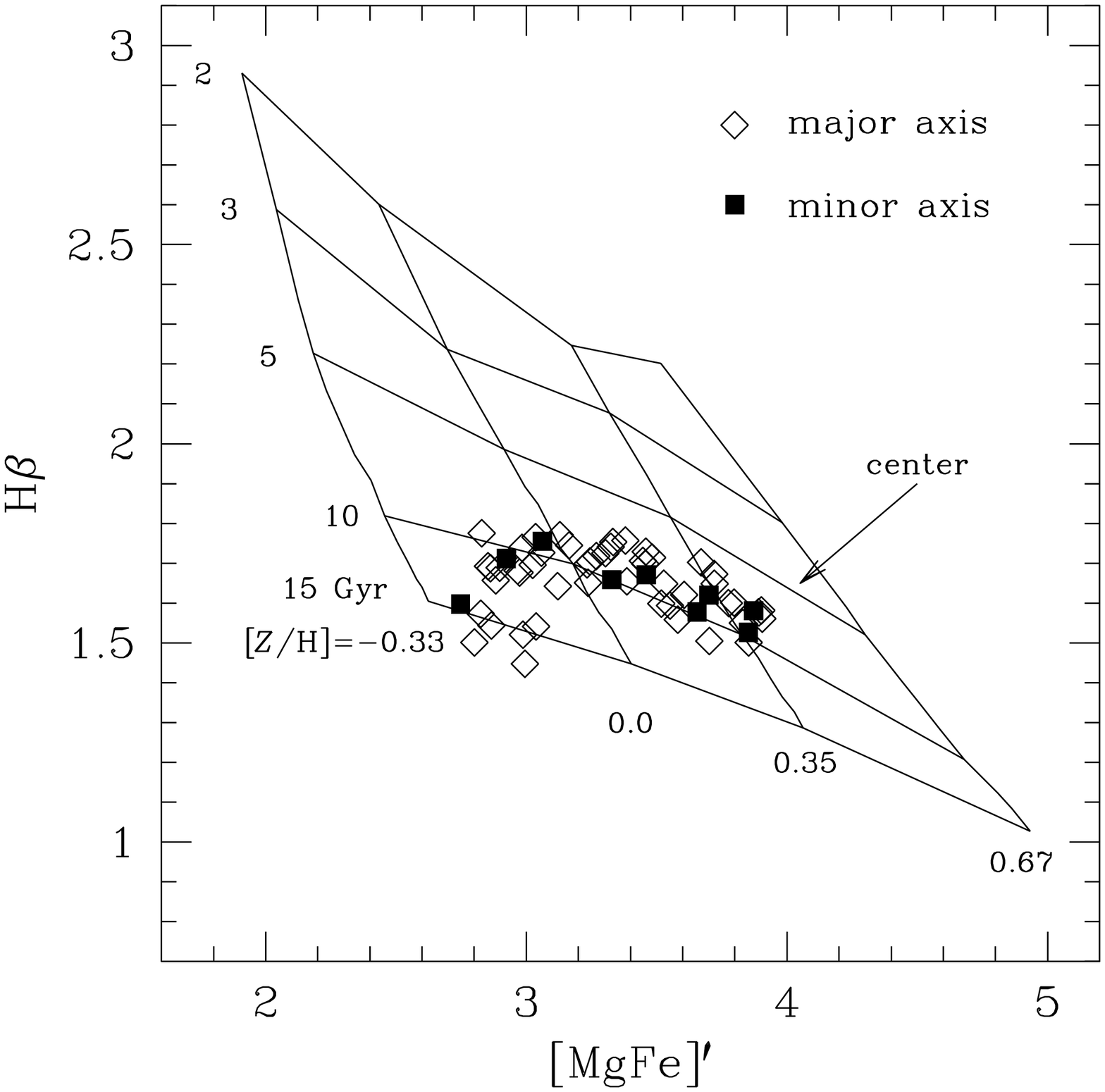}{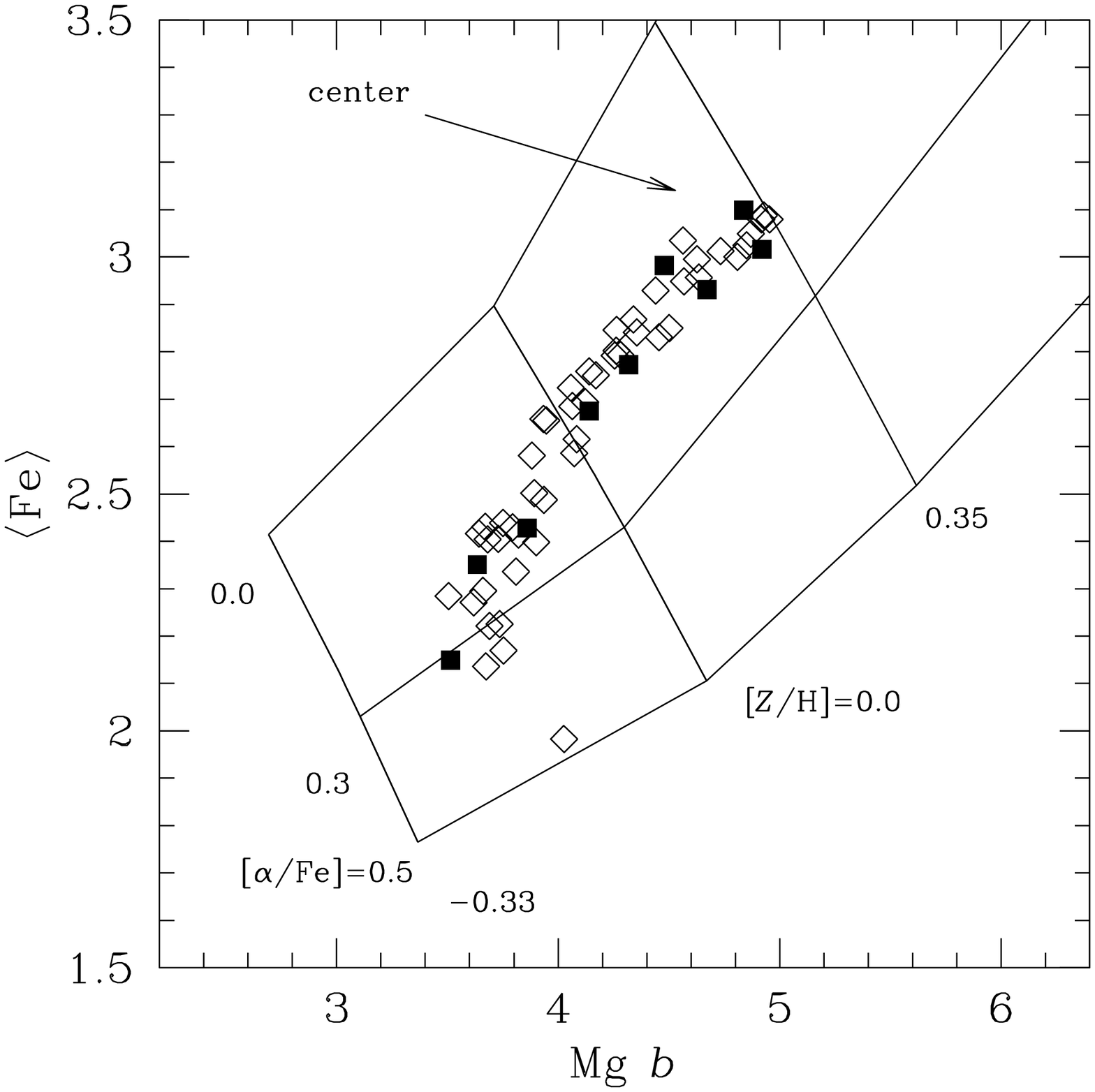}
\caption{Lick absorption line indices \MgFe\ vs.\ \Hb\
(left panel) and \Mgb\ vs.\ $\Fe=0.5 ({\rm Fe5270}+{\rm Fe5335})$
(right panel).  Overplotted are the TMB03 stellar population models
for various ages, metallicities, and \aFe\ ratios as indicated by the
labels in the diagrams. In the left panel, \aFe\ is solar, while in
the right panel the age is fixed to $12\;$Gyr. Open diamonds and filled
squares are along the major and minor axes, respectively.}
\end{figure}

\begin{figure}
\figurenum{3}
\epsscale{1.0}
\plotone{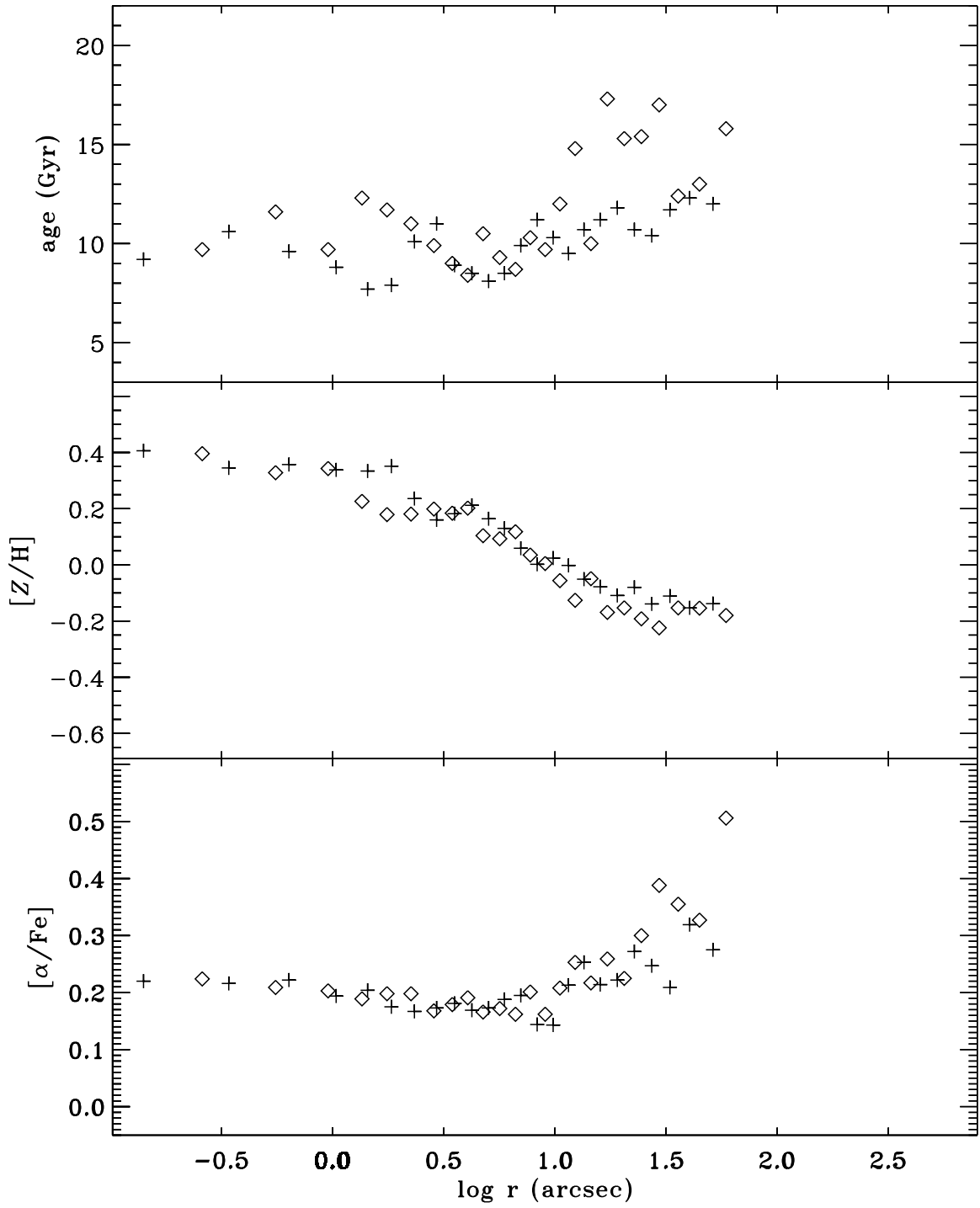}
\caption{Stellar population parameters derived from Fig. 2 as
functions of galactic radius. The effective radius is at $\sim
95\;$arcsec. Diamonds and plus signs are positive and negative
radial directions, respectively, along the major axis.}
\end{figure}

\begin{figure}
\figurenum{4}
\epsscale{1.0}
\plotone{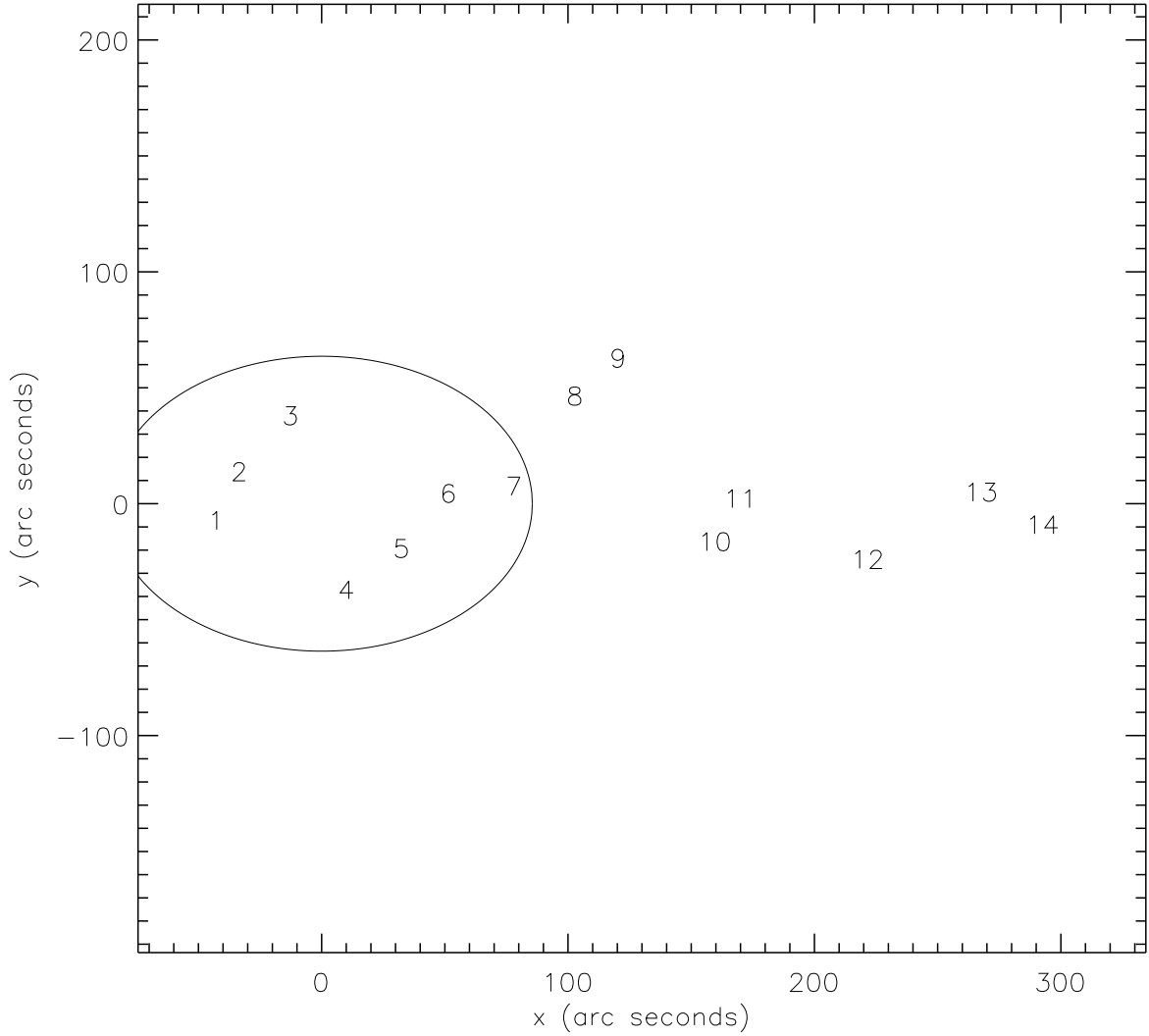}
\caption{Positions of the 14 bright PNs relative to the core 
of NGC 4697. The area and orientation of this figure are as in Fig. 7 of
M\'endez et al. (2001). The ellipse is a schematic representation of
NGC 4697. The origin of coordinates is located at the light center of
the galaxy. The effective radius of NGC 4697 is 95 arc seconds.}
\end{figure}

\begin{figure}
\figurenum{5}
\epsscale{0.6}
\plotone{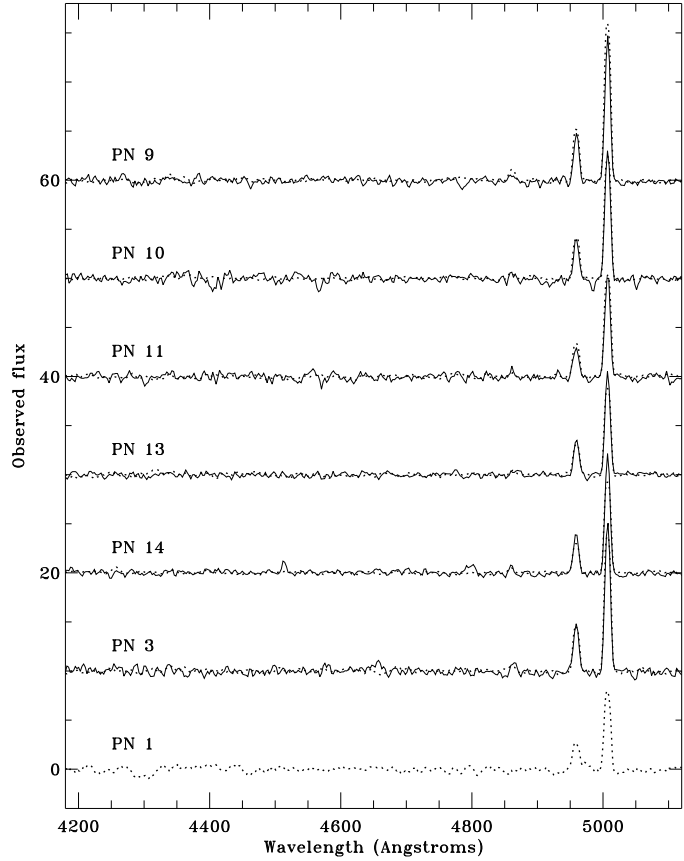}
\caption{Comparison of individual PN spectra taken with FORS1
(full lines) and LRIS-B (dotted lines). The PNs are designated with 
the numbers shown in Fig. 4. The fluxes are expressed in units of
10$^{-18}$ erg cm$^{-2}$ s$^{-1}$ \AA$^{-1}$. The levels of zero
intensity are separated by 10 flux units for easier comparison.
The higher-resolution LRIS-B spectra have been smoothed with a boxcar=7.
PN 1 could not be observed with FORS1.}
\end{figure}

\begin{figure}
\figurenum{6}
\epsscale{0.6}
\plotone{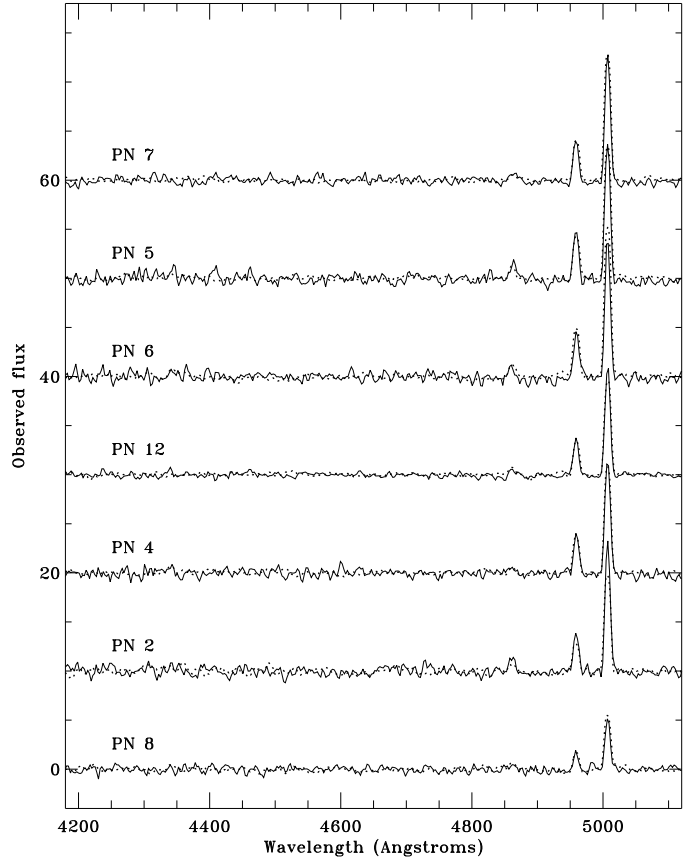}
\caption{Comparison of individual PN spectra taken with FORS1
(full lines) and LRIS-B (dotted lines). The PNs are designated with 
the numbers shown in Fig. 4. The fluxes are expressed in units of
10$^{-18}$ erg cm$^{-2}$ s$^{-1}$ \AA$^{-1}$. The levels of zero
intensity are separated by 10 flux units for easier comparison.
The higher-resolution LRIS-B spectra have been smoothed with a boxcar=7.}
\end{figure}

\begin{figure}
\figurenum{7}
\epsscale{0.6}
\plotone{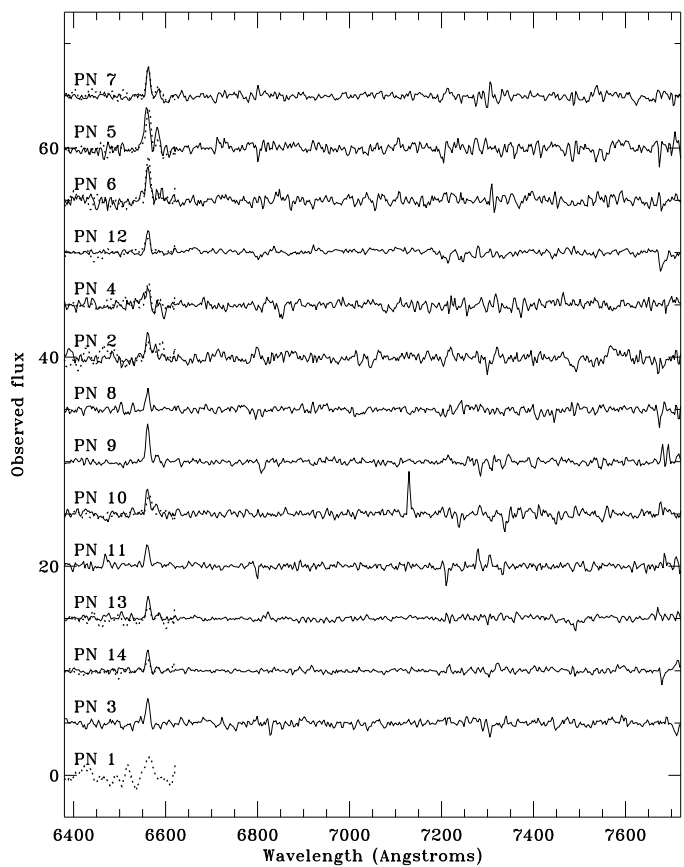}
\caption{Comparison of individual PN spectra taken with FORS1
(full lines) and LRIS-B (dotted lines). The PNs are designated with 
the numbers shown in Fig. 4. The fluxes are expressed in units of
10$^{-18}$ erg cm$^{-2}$ s$^{-1}$ \AA$^{-1}$. The levels of zero
intensity are separated by 5 flux units for easier comparison.
The higher-resolution LRIS-B spectra have been smoothed with a boxcar=7.
The feature longward of H$\alpha$ 6563 is not [N II] $\lambda$6584; it is 
due to residual sky emission. The sky emissions are blueshifted because
the PN emission lines have been shifted to their rest values.}
\end{figure}

\begin{figure}
\figurenum{8}
\epsscale{0.6}
\plotone{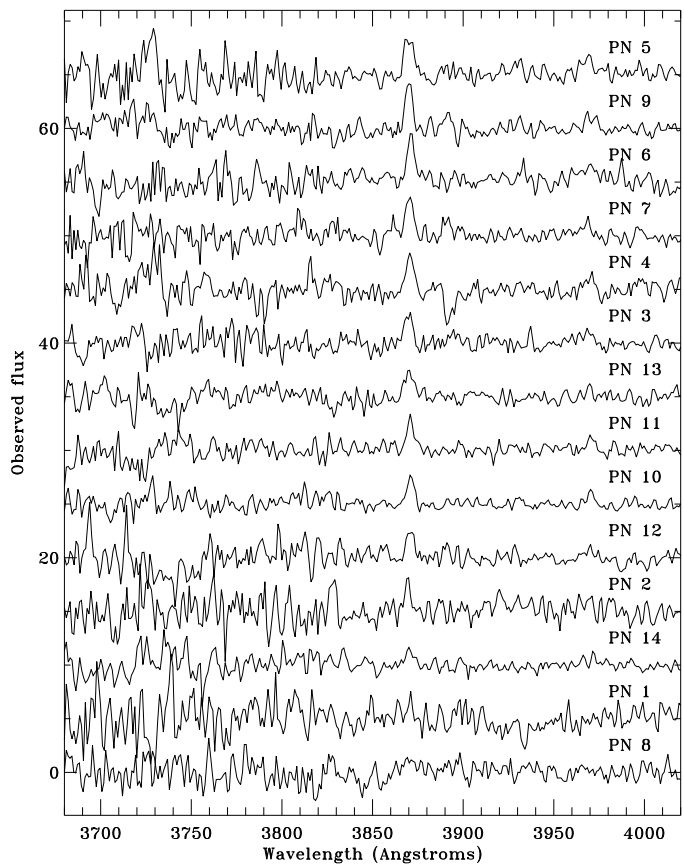}
\caption{Individual LRIS-B PN spectra from 3700 to 4000 \AA.
The fluxes are expressed in units of 
10$^{-18}$ erg cm$^{-2}$ s$^{-1}$ \AA$^{-1}$. The levels of zero
intensity are separated by 5 flux units for easier comparison.
The visible features are [Ne III] $\lambda$3868 and (marginally) 
[Ne III] $\lambda$3967. PN 5 may show some [O II] $\lambda$3727.}
\end{figure}

\begin{figure}
\figurenum{9}
\epsscale{1.0}
\plotone{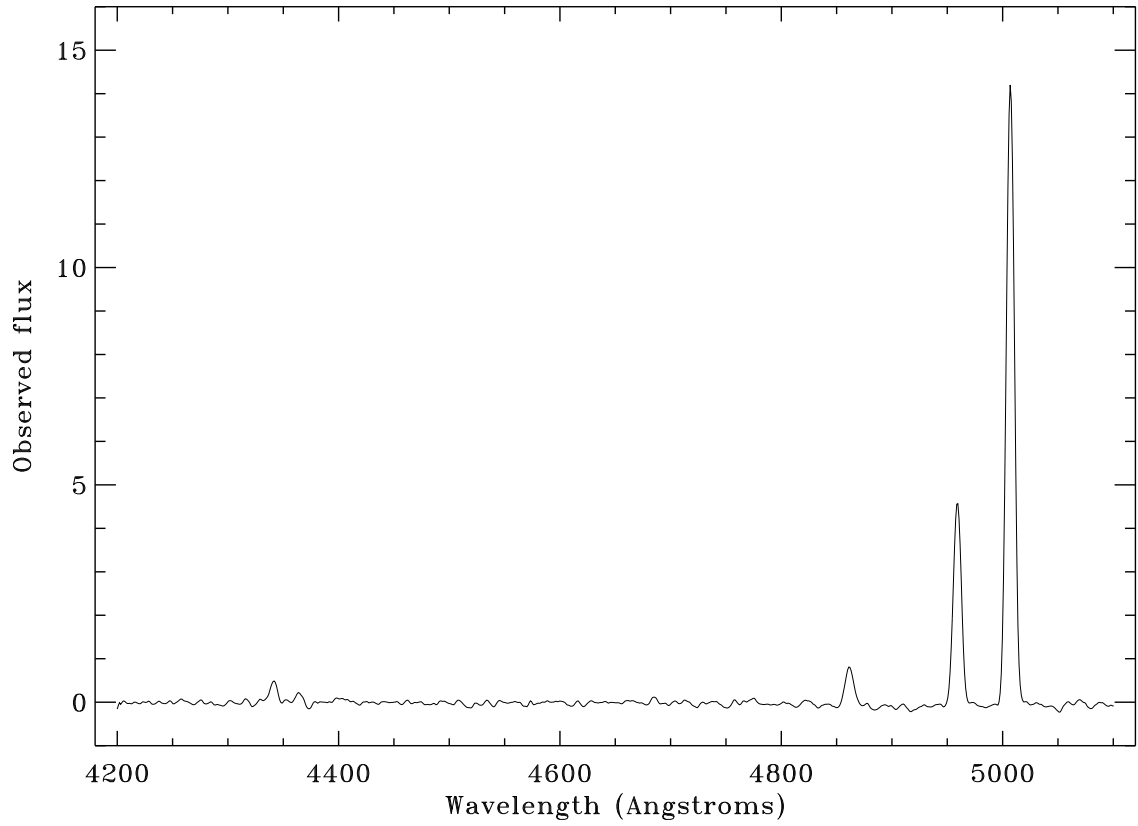}
\caption{Average of all spectra of the 14 PNs in NGC 4697.
Visible features are H$\gamma$ $\lambda$4340, [O III] $\lambda$4363,
possibly a very weak He II $\lambda$4686, H$\beta$ $\lambda$4861, [O III]
$\lambda\lambda$4959, 5007.}
\end{figure}

\begin{figure}
\figurenum{10}
\epsscale{1.0}
\plotone{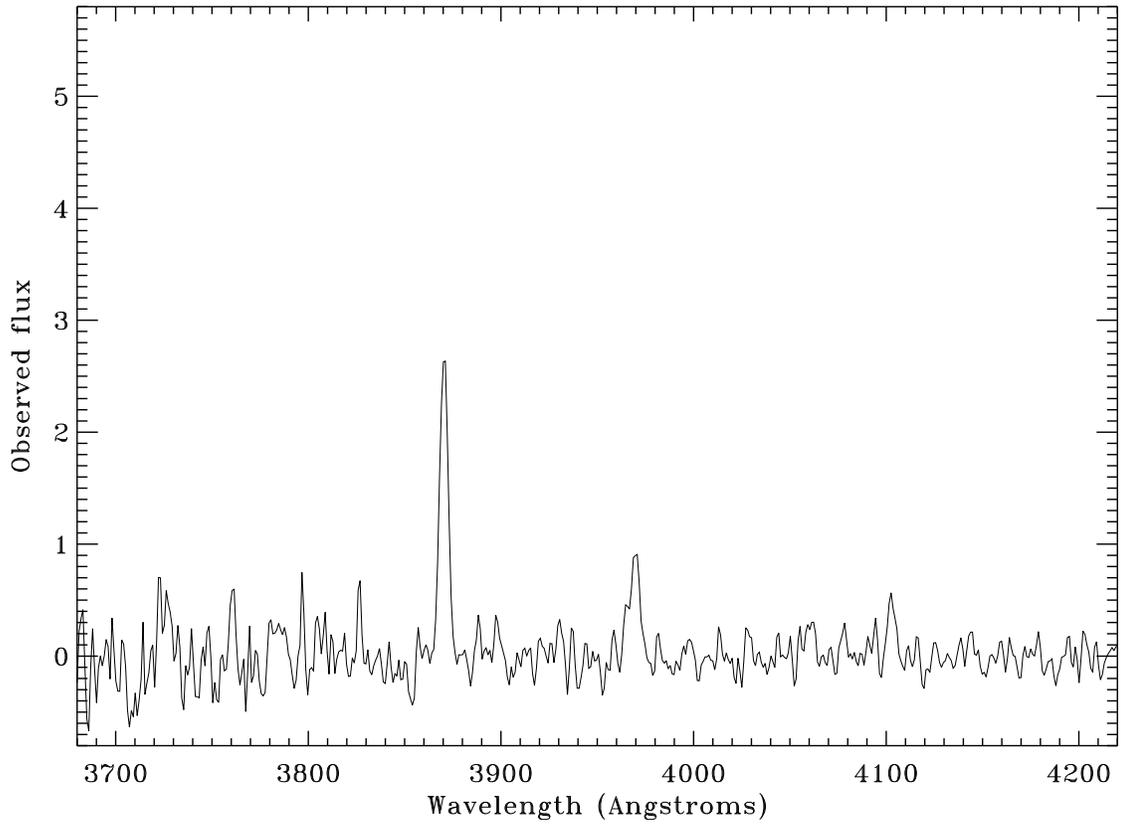}
\caption{Average of all LRIS-B spectra of the 14 PNs in NGC 4697.
Visible features are [Ne III] $\lambda\lambda$3868, 3967,
H$\delta$ $\lambda$4101.}
\end{figure}

\begin{figure}
\figurenum{11}
\epsscale{1.0}
\plotone{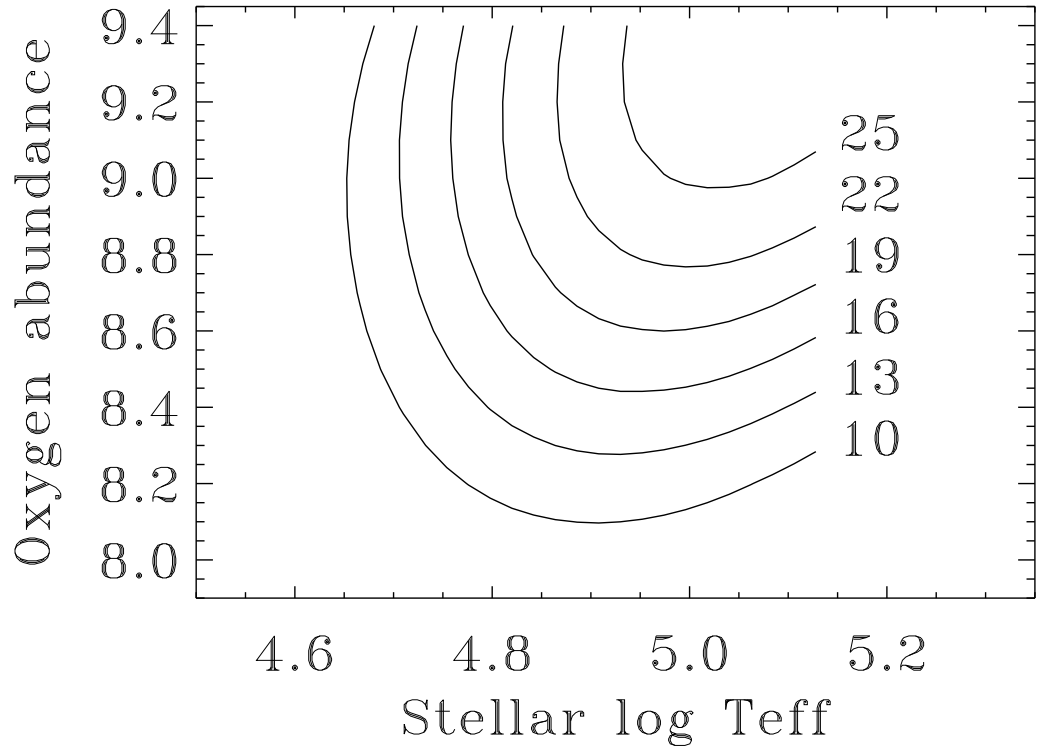}
\caption{This is our imitation of Dopita et al.'s Fig. 3 (see text).
Our grid was calculated using CLOUDY. 
The contours give the 5007/$H\beta$ ratio as a function of stellar 
effective temperature (we assume blackbody energy distributions) and
oxygen abundance (logarithmic, in the scale where H=12).}
\end{figure}

\begin{figure}
\figurenum{12}
\epsscale{1.0}
\plotone{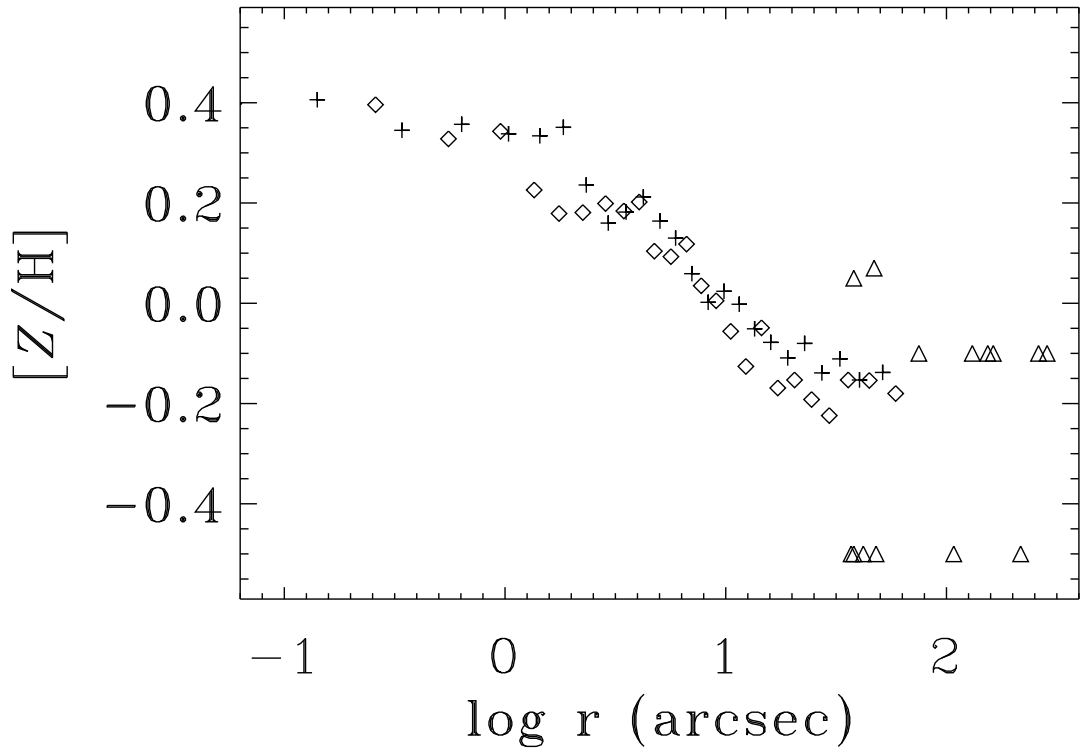}
\caption{The combined information about metallicity from 
absorption-line data (diamonds and plus signs as in Fig. 3) 
and from the PNs (triangles pointing upward indicate lower 
limits to the abundances) plotted as a function
of projected distance to the center of NGC 4697. One effective radius
is at $\log r = 1.98$.}
\end{figure}

\begin{figure}
\figurenum{13}
\plottwo{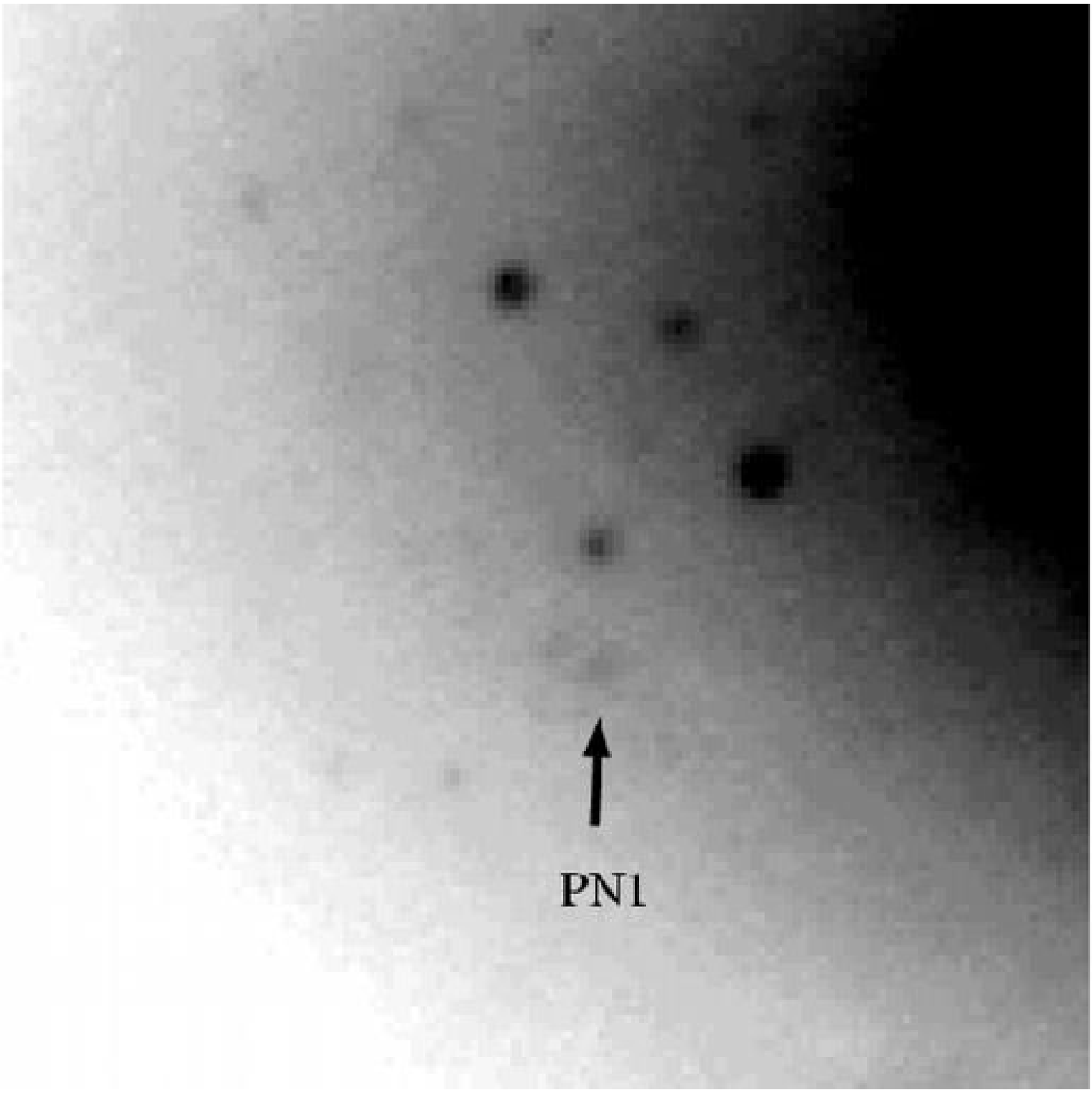}{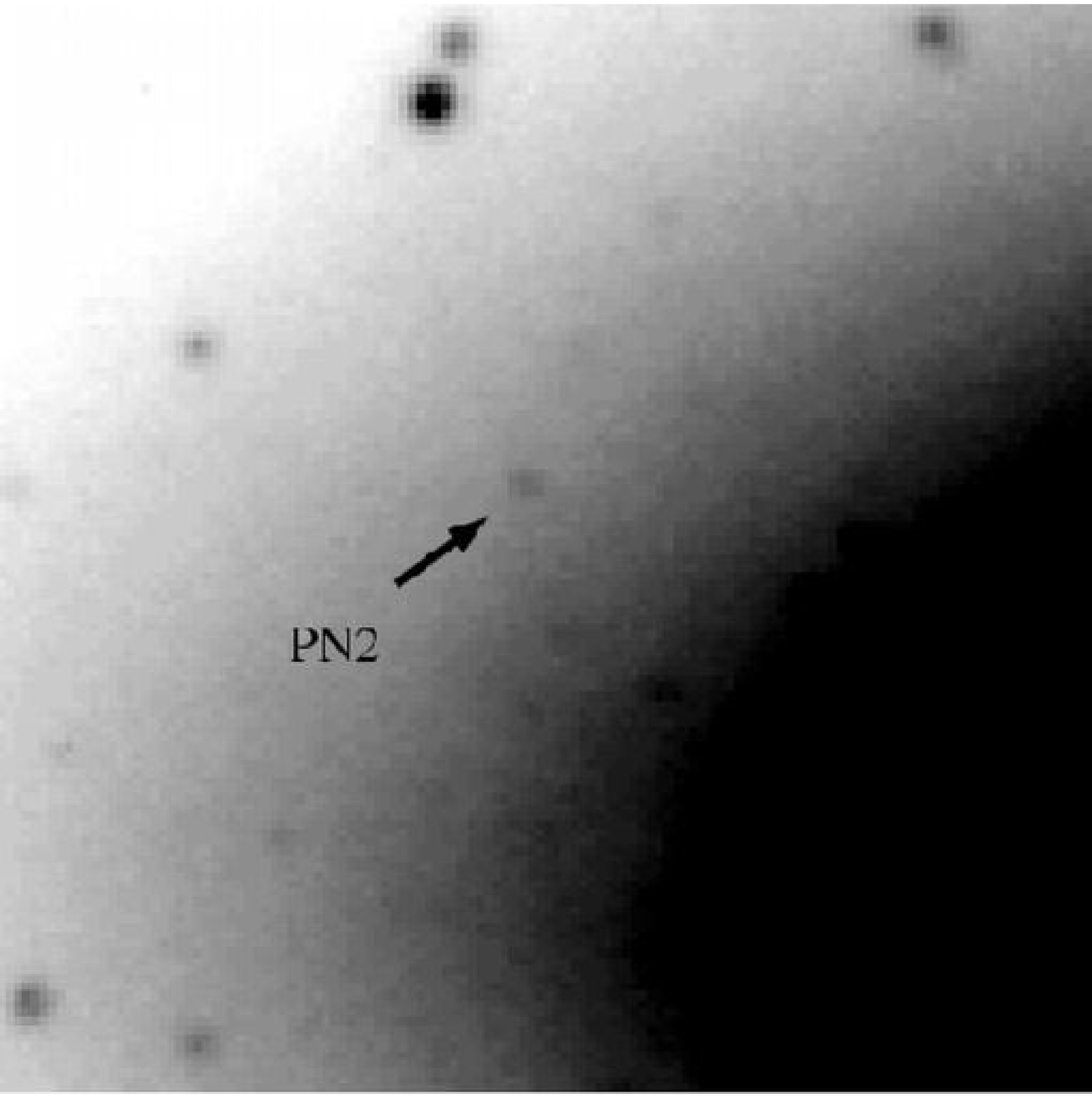}
\plottwo{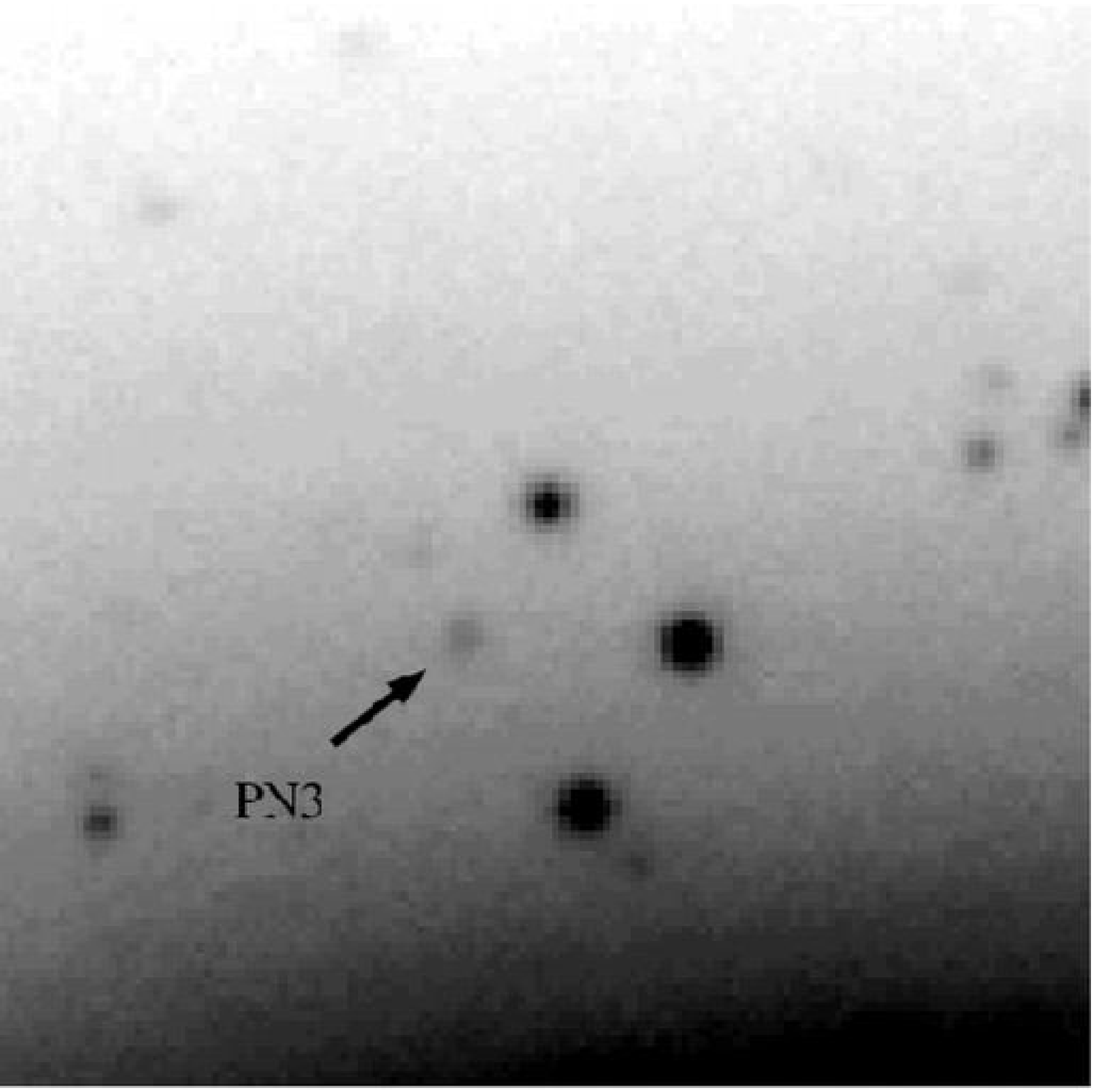}{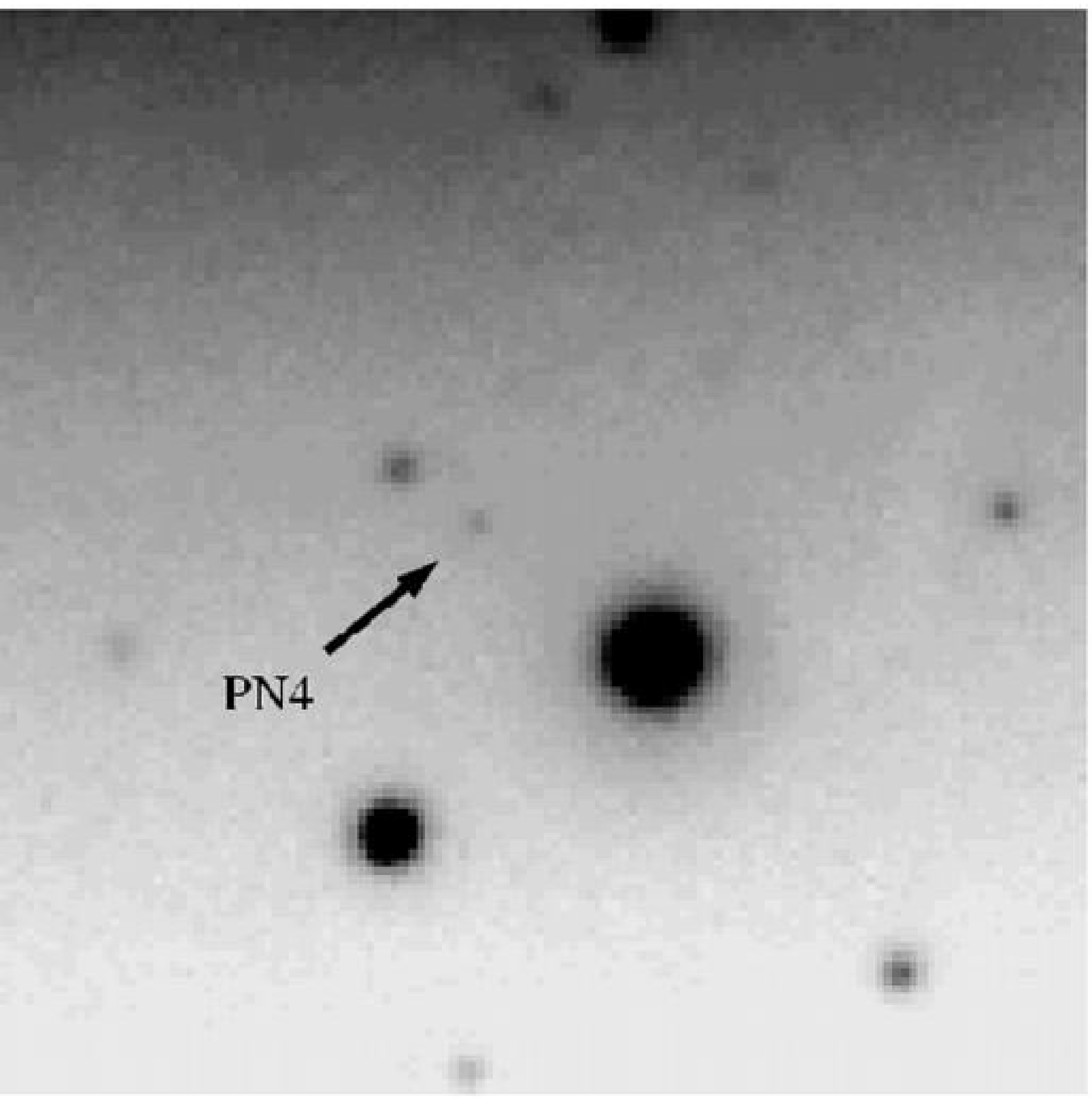}
\plottwo{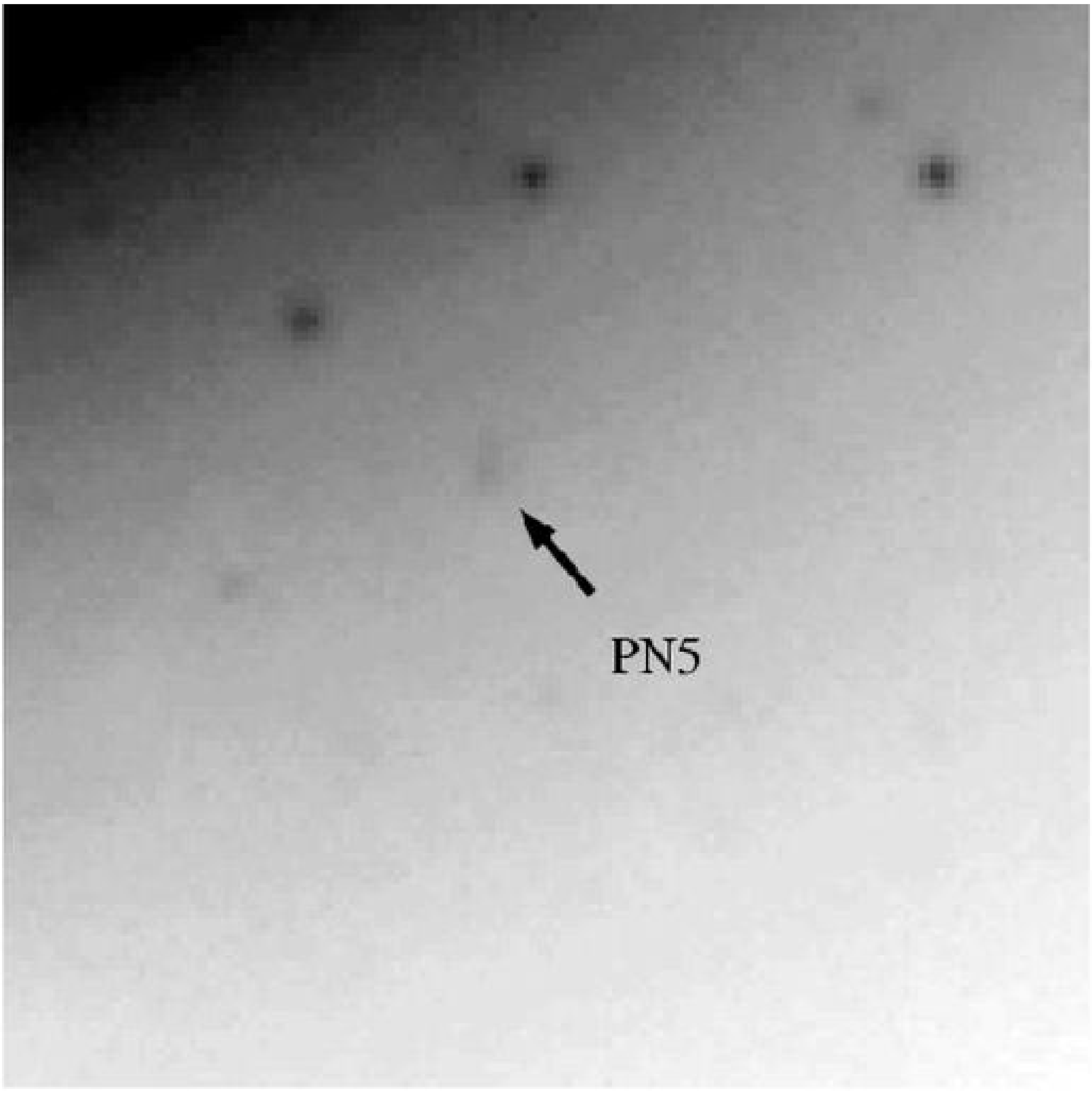}{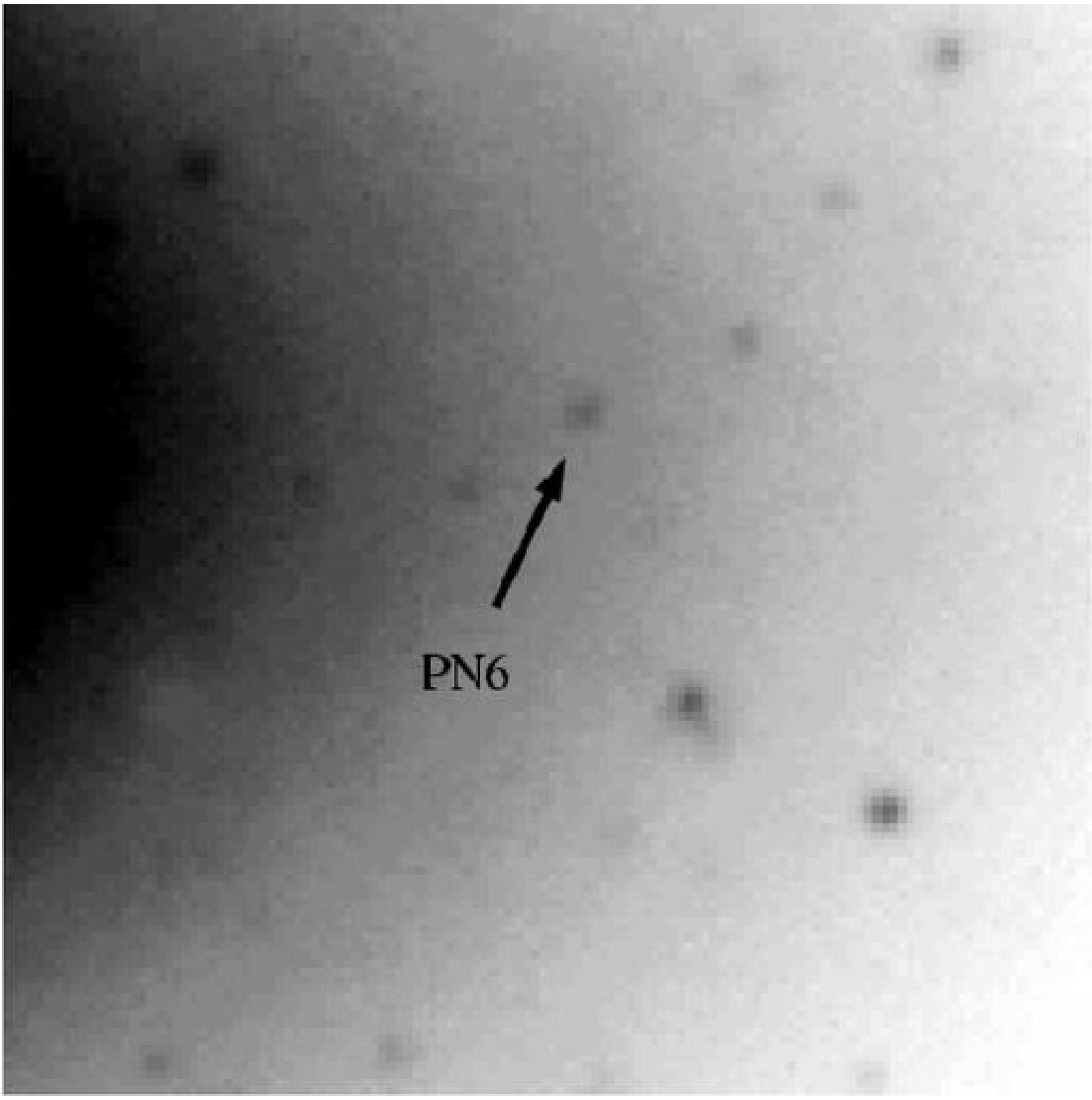}
\caption{Finding charts for the PNs studied in this work.}
\end{figure}

\begin{figure}
\figurenum{14}
\plottwo{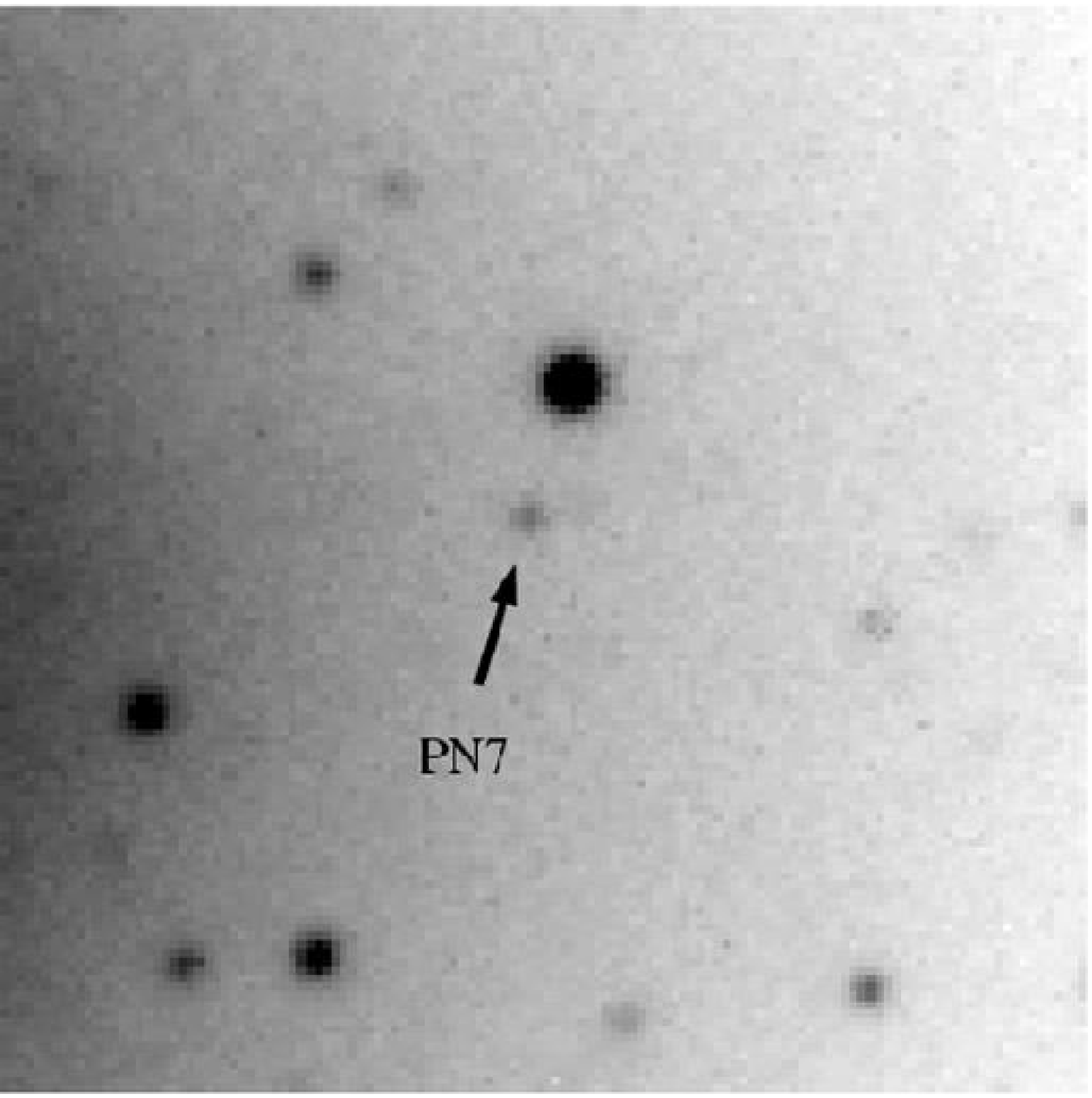}{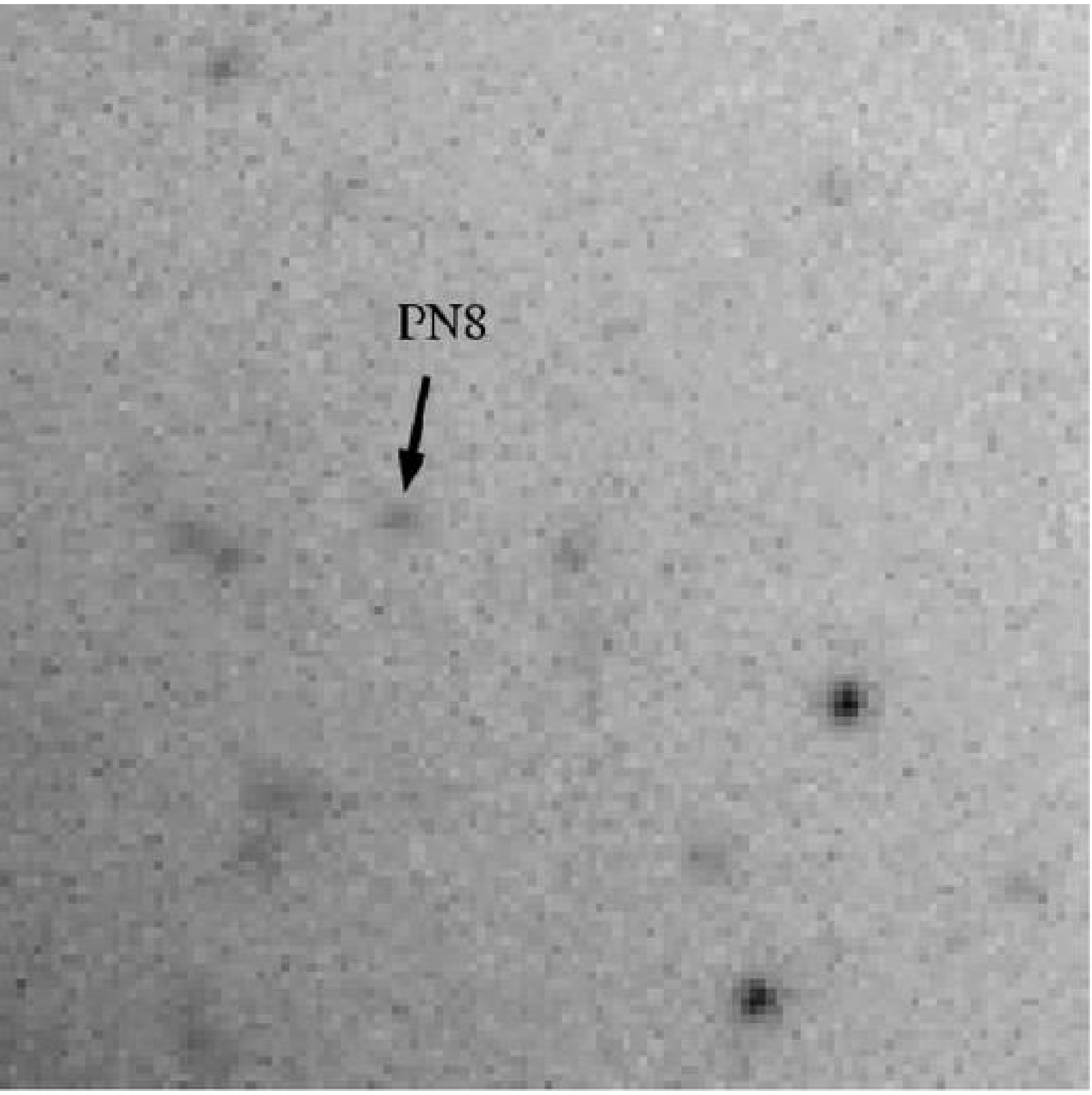}
\plottwo{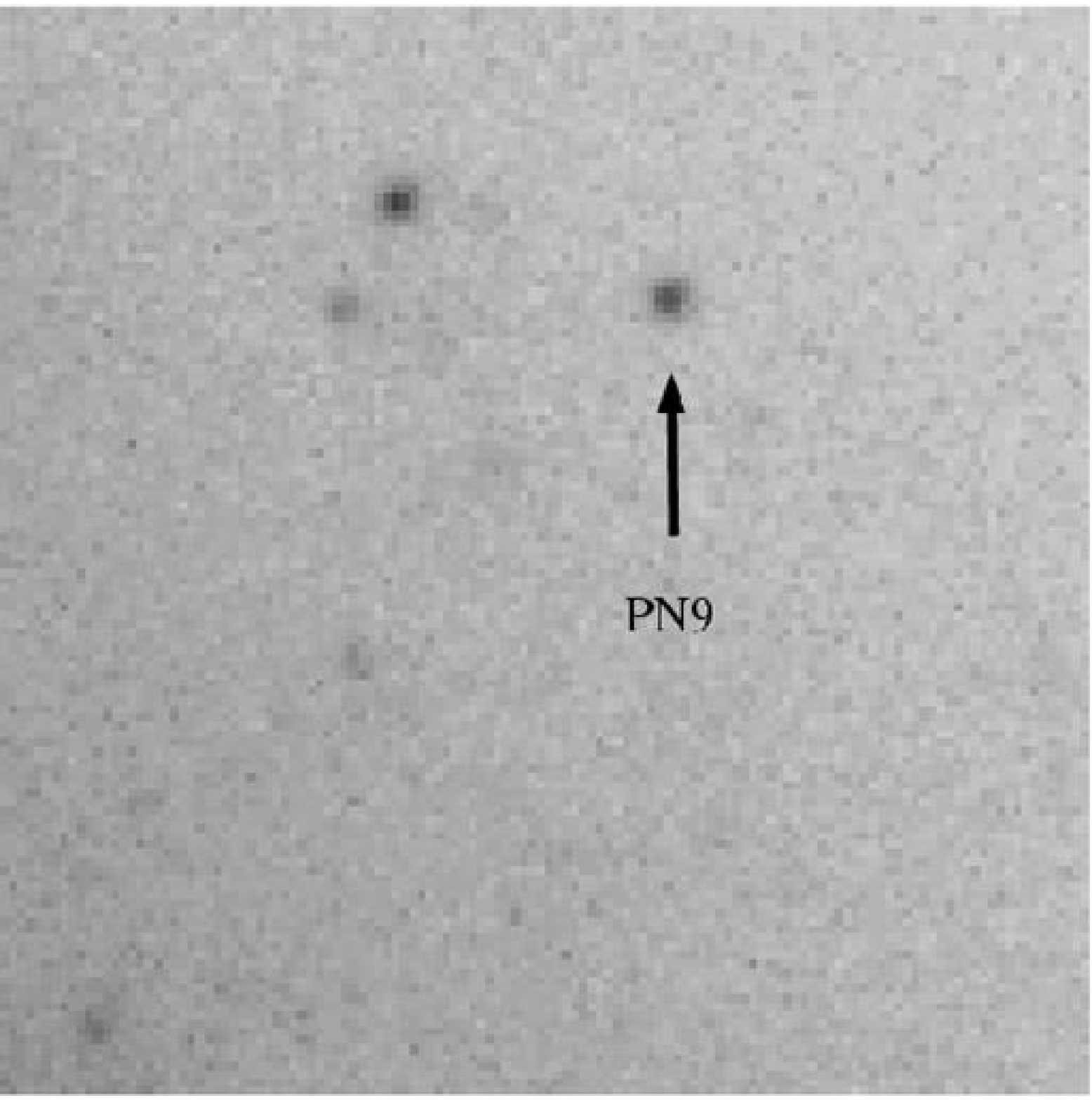}{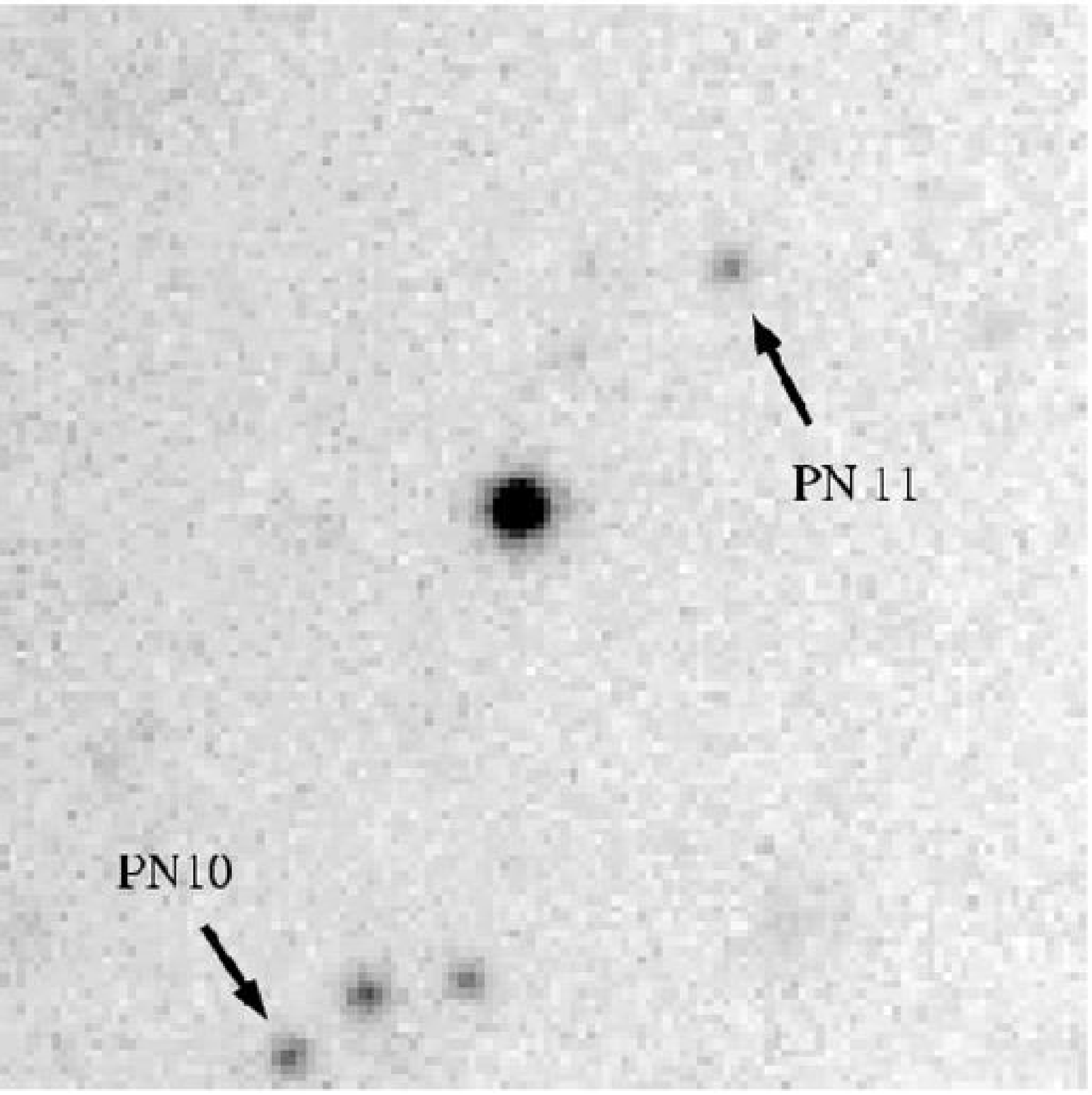}
\plottwo{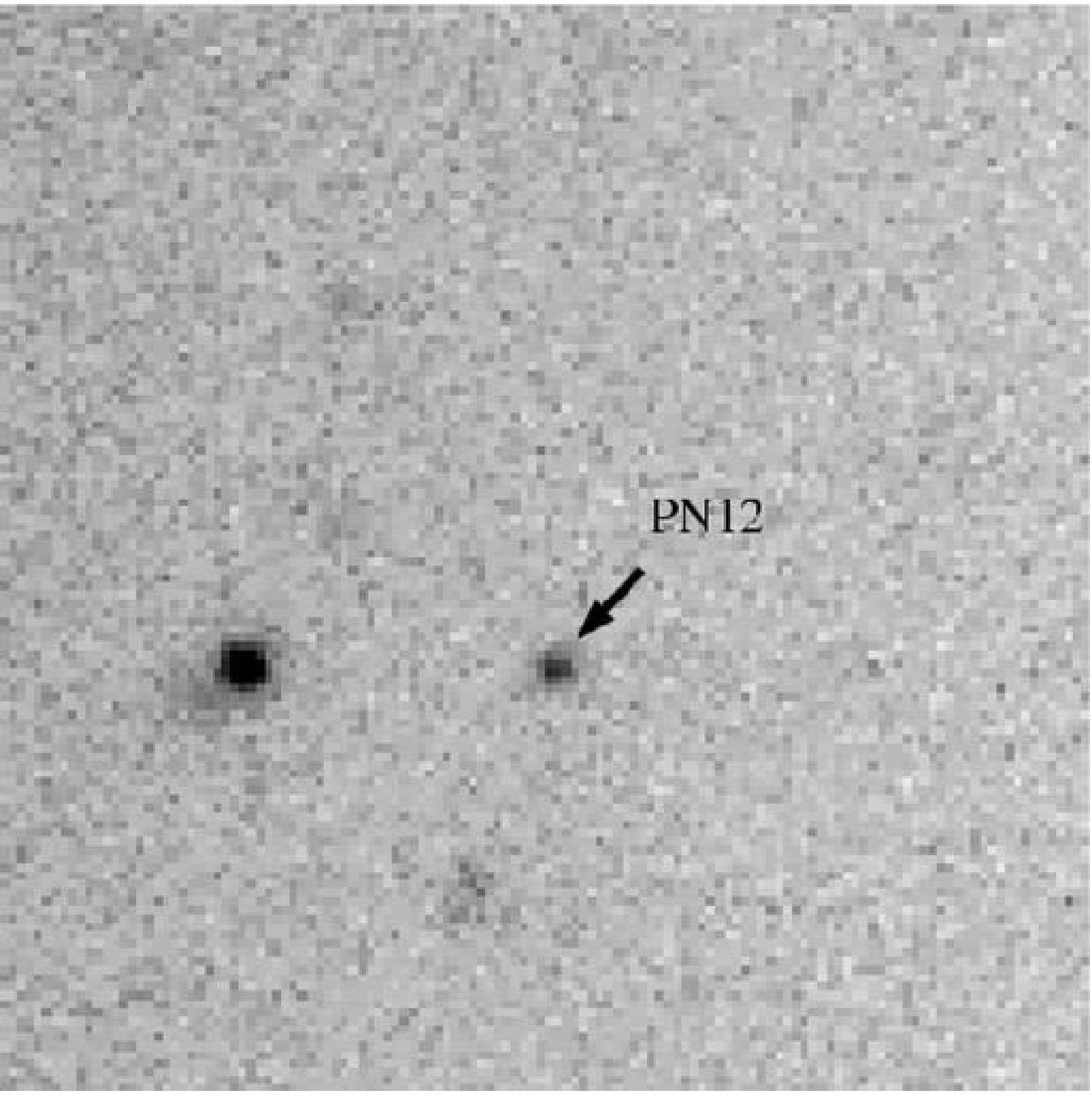}{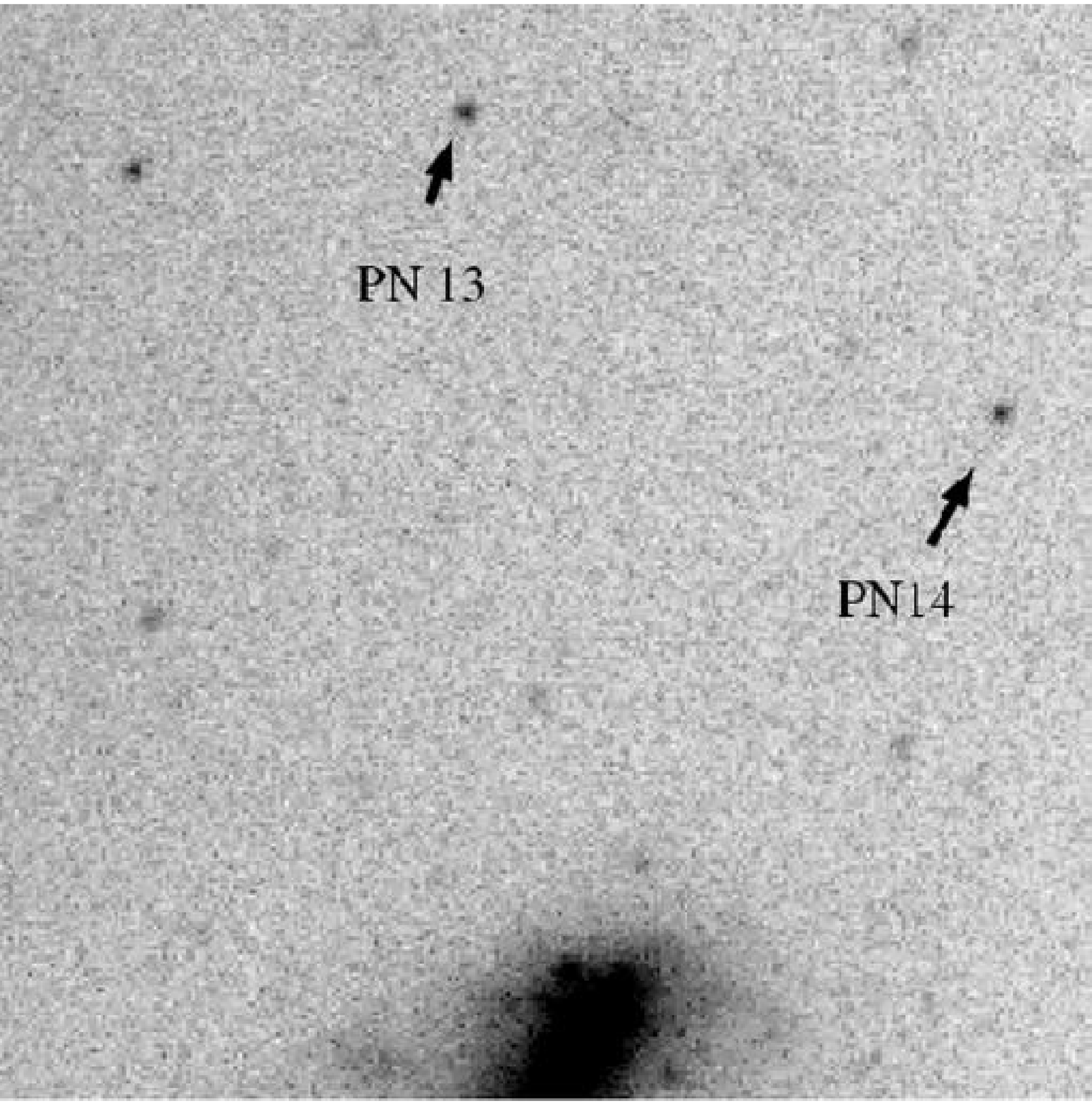}
\caption{Finding charts for the PNs studied in this work.}
\end{figure}

\begin{deluxetable}{rrrrrrrrrrrrr}
\tablecaption{Integrated stellar populations in NGC 4697 
(major axis)\tablenotemark{a} \label{tbl-1}}
\tablewidth{0pt}
\rotate
\tabletypesize{\tiny}
\tablehead{
\colhead{Radius (arcsec)} & \colhead{\Hb} & \colhead{d\Hb} &
\colhead{\Mgb} & \colhead{d\Mgb} & \colhead{\Fe} & \colhead{d\Fe} & 
\colhead{Age (Gyr)} & \colhead{dAge} &
\colhead{[\ZH]} & \colhead{d[\ZH]} & \colhead{[\aFe]} & \colhead{d[\aFe]} } 
\startdata

 -51.380 & 1.689 & 0.095 & 3.661 & 0.107 & 2.295 & 0.142 & 12.0 & 2.0 & -0.138 & 0.044 & 0.275 & 0.064 \\
 -40.370 & 1.688 & 0.065 & 3.691 & 0.073 & 2.221 & 0.099 & 12.3 & 1.4 & -0.153 & 0.034 & 0.319 & 0.052 \\
 -32.900 & 1.682 & 0.059 & 3.642 & 0.067 & 2.416 & 0.092 & 11.7 & 1.2 & -0.111 & 0.032 & 0.209 & 0.037 \\
 -27.250 & 1.775 & 0.062 & 3.506 & 0.071 & 2.284 & 0.089 & 10.4 & 1.0 & -0.139 & 0.031 & 0.247 & 0.046 \\
 -22.770 & 1.739 & 0.053 & 3.810 & 0.060 & 2.336 & 0.076 & 10.7 & 1.0 & -0.080 & 0.029 & 0.272 & 0.040 \\
 -19.080 & 1.676 & 0.039 & 3.682 & 0.043 & 2.404 & 0.061 & 11.8 & 1.1 & -0.109 & 0.029 & 0.222 & 0.033 \\
 -15.990 & 1.696 & 0.029 & 3.752 & 0.032 & 2.439 & 0.045 & 11.2 & 0.6 & -0.078 & 0.017 & 0.214 & 0.022 \\
 -13.510 & 1.726 & 0.028 & 3.902 & 0.031 & 2.398 & 0.044 & 10.7 & 0.5 & -0.051 & 0.017 & 0.253 & 0.023 \\
 -11.510 & 1.770 & 0.025 & 3.936 & 0.028 & 2.487 & 0.040 &  9.5 & 0.6 & -0.002 & 0.023 & 0.213 & 0.018 \\
  -9.817 & 1.697 & 0.019 & 3.933 & 0.021 & 2.658 & 0.030 & 10.3 & 0.4 &  0.024 & 0.015 & 0.143 & 0.011 \\
  -8.321 & 1.651 & 0.021 & 3.946 & 0.024 & 2.655 & 0.034 & 11.2 & 0.4 &  0.002 & 0.017 & 0.144 & 0.013 \\
  -7.023 & 1.718 & 0.016 & 4.084 & 0.018 & 2.615 & 0.026 &  9.9 & 0.3 &  0.059 & 0.015 & 0.195 & 0.009 \\
  -5.928 & 1.753 & 0.013 & 4.123 & 0.015 & 2.694 & 0.021 &  8.5 & 0.3 &  0.130 & 0.014 & 0.188 & 0.008 \\
  -5.032 & 1.757 & 0.016 & 4.140 & 0.018 & 2.759 & 0.026 &  8.1 & 0.4 &  0.164 & 0.015 & 0.173 & 0.009 \\
  -4.231 & 1.714 & 0.014 & 4.264 & 0.016 & 2.846 & 0.023 &  8.5 & 0.3 &  0.212 & 0.014 & 0.169 & 0.008 \\
  -3.536 & 1.706 & 0.019 & 4.256 & 0.021 & 2.792 & 0.030 &  8.9 & 0.5 &  0.182 & 0.019 & 0.181 & 0.008 \\
  -2.936 & 1.599 & 0.015 & 4.355 & 0.017 & 2.841 & 0.025 & 11.0 & 0.4 &  0.160 & 0.016 & 0.173 & 0.009 \\
  -2.334 & 1.620 & 0.012 & 4.439 & 0.013 & 2.929 & 0.020 & 10.1 & 0.3 &  0.236 & 0.014 & 0.167 & 0.008 \\
  -1.838 & 1.673 & 0.014 & 4.564 & 0.017 & 3.035 & 0.023 &  7.9 & 0.4 &  0.351 & 0.012 & 0.175 & 0.007 \\
  -1.437 & 1.702 & 0.012 & 4.569 & 0.014 & 2.949 & 0.019 &  7.7 & 0.3 &  0.334 & 0.012 & 0.204 & 0.006 \\
  -1.037 & 1.649 & 0.012 & 4.626 & 0.014 & 2.995 & 0.019 &  8.8 & 0.4 &  0.338 & 0.012 & 0.194 & 0.007 \\
  -0.636 & 1.601 & 0.013 & 4.809 & 0.014 & 3.000 & 0.016 &  9.6 & 0.3 &  0.357 & 0.009 & 0.222 & 0.005 \\
  -0.341 & 1.550 & 0.015 & 4.850 & 0.015 & 3.025 & 0.018 & 10.6 & 0.4 &  0.345 & 0.013 & 0.216 & 0.006 \\
  -0.141 & 1.581 & 0.017 & 4.929 & 0.022 & 3.088 & 0.024 &  9.2 & 0.5 &  0.406 & 0.014 & 0.220 & 0.006 \\
   0.059 & 1.577 & 0.016 & 4.917 & 0.020 & 3.080 & 0.022 &  9.4 & 0.4 &  0.398 & 0.014 & 0.219 & 0.007 \\
   0.259 & 1.561 & 0.014 & 4.954 & 0.016 & 3.080 & 0.018 &  9.7 & 0.4 &  0.396 & 0.011 & 0.224 & 0.006 \\
   0.554 & 1.501 & 0.018 & 4.870 & 0.015 & 3.049 & 0.018 & 11.6 & 0.5 &  0.328 & 0.017 & 0.209 & 0.006 \\
   0.954 & 1.600 & 0.018 & 4.734 & 0.015 & 3.012 & 0.018 &  9.7 & 0.4 &  0.343 & 0.014 & 0.203 & 0.007 \\
   1.355 & 1.505 & 0.016 & 4.635 & 0.017 & 2.957 & 0.022 & 12.3 & 0.4 &  0.226 & 0.015 & 0.189 & 0.007 \\
   1.756 & 1.558 & 0.018 & 4.502 & 0.020 & 2.850 & 0.026 & 11.7 & 0.5 &  0.179 & 0.018 & 0.198 & 0.011 \\
   2.252 & 1.594 & 0.016 & 4.456 & 0.017 & 2.831 & 0.022 & 11.0 & 0.4 &  0.181 & 0.015 & 0.198 & 0.007 \\
   2.854 & 1.648 & 0.015 & 4.341 & 0.020 & 2.868 & 0.021 &  9.9 & 0.4 &  0.199 & 0.018 & 0.168 & 0.008 \\
   3.454 & 1.699 & 0.016 & 4.263 & 0.021 & 2.802 & 0.023 &  9.0 & 0.4 &  0.184 & 0.014 & 0.179 & 0.009 \\
   4.053 & 1.729 & 0.020 & 4.282 & 0.026 & 2.793 & 0.028 &  8.4 & 0.5 &  0.202 & 0.018 & 0.191 & 0.011 \\
   4.749 & 1.655 & 0.019 & 4.171 & 0.025 & 2.750 & 0.027 & 10.5 & 0.4 &  0.104 & 0.018 & 0.166 & 0.012 \\
   5.645 & 1.725 & 0.016 & 4.064 & 0.021 & 2.685 & 0.022 &  9.3 & 0.5 &  0.093 & 0.017 & 0.172 & 0.008 \\
   6.646 & 1.741 & 0.024 & 4.058 & 0.033 & 2.724 & 0.035 &  8.7 & 0.4 &  0.118 & 0.022 & 0.162 & 0.012 \\
   7.743 & 1.704 & 0.018 & 4.072 & 0.024 & 2.586 & 0.026 & 10.3 & 0.3 &  0.035 & 0.016 & 0.201 & 0.011 \\
   9.040 & 1.745 & 0.025 & 3.881 & 0.033 & 2.581 & 0.036 &  9.7 & 0.7 &  0.005 & 0.025 & 0.162 & 0.015 \\
  10.530 & 1.643 & 0.031 & 3.892 & 0.041 & 2.502 & 0.044 & 12.0 & 0.7 & -0.056 & 0.020 & 0.208 & 0.024 \\
  12.330 & 1.541 & 0.033 & 3.821 & 0.045 & 2.414 & 0.047 & 14.8 & 0.8 & -0.126 & 0.017 & 0.253 & 0.023 \\
  14.520 & 1.765 & 0.036 & 3.794 & 0.049 & 2.430 & 0.052 & 10.0 & 0.8 & -0.049 & 0.024 & 0.217 & 0.032 \\
  17.210 & 1.447 & 0.041 & 3.729 & 0.056 & 2.405 & 0.059 & 17.3 & 0.9 & -0.169 & 0.018 & 0.259 & 0.025 \\
  20.500 & 1.520 & 0.051 & 3.670 & 0.069 & 2.431 & 0.072 & 15.3 & 1.3 & -0.153 & 0.023 & 0.225 & 0.044 \\
  24.480 & 1.544 & 0.065 & 3.618 & 0.088 & 2.271 & 0.091 & 15.4 & 1.6 & -0.192 & 0.024 & 0.300 & 0.047 \\
  29.450 & 1.501 & 0.089 & 3.674 & 0.092 & 2.136 & 0.095 & 17.0 & 2.0 & -0.224 & 0.036 & 0.388 & 0.037 \\
  35.860 & 1.692 & 0.103 & 3.753 & 0.106 & 2.170 & 0.108 & 12.4 & 2.0 & -0.153 & 0.049 & 0.355 & 0.056 \\
  44.710 & 1.657 & 0.155 & 3.735 & 0.157 & 2.225 & 0.155 & 13.0 & 2.8 & -0.154 & 0.070 & 0.327 & 0.086 \\
  58.860 & 1.573 & 0.243 & 4.026 & 0.249 & 1.983 & 0.242 & 15.8 & 3.5 & -0.180 & 0.062 & 0.506 & 0.099 \\
\enddata 
\tablenotetext{a}{The radii in column 1 are measured from 
the center of the galaxy and defined positive towards NE. 
Stellar population parameters (cols. 8, 10, 12) and their uncertainties 
(cols. 9, 11, 13) are derived from the absorption line indices 
(cols. 2--7) using the stellar population models of TMB03.}
\end{deluxetable}

\begin{deluxetable}{rrrrrrrrrrrrr}
\tablecaption{Integrated stellar populations in NGC 4697 
(minor axis)\tablenotemark{a} \label{tbl-2}}
\tablewidth{0pt}
\rotate
\tabletypesize{\tiny}
\tablehead{
\colhead{Radius (arcsec)} & \colhead{\Hb} & \colhead{d\Hb} &
\colhead{\Mgb} & \colhead{d\Mgb} & \colhead{\Fe} & \colhead{d\Fe} & 
\colhead{Age (Gyr)} & \colhead{dAge} &
\colhead{[\ZH]} & \colhead{d[\ZH]} & \colhead{[\aFe]} & \colhead{d[\aFe]} } 
\startdata

 -7.398 & 1.712 & 0.034 & 3.635 & 0.033 & 2.351 & 0.040 & 11.3 & 0.7 & -0.119 & 0.018 & 0.239 & 0.018 \\
 -3.230 & 1.659 & 0.023 & 4.140 & 0.024 & 2.675 & 0.028 & 10.7 & 0.5 &  0.067 & 0.021 & 0.183 & 0.014 \\
 -1.444 & 1.578 & 0.023 & 4.479 & 0.023 & 2.982 & 0.028 & 10.6 & 0.6 &  0.243 & 0.020 & 0.156 & 0.008 \\
 -0.486 & 1.581 & 0.020 & 4.837 & 0.020 & 3.099 & 0.024 &  9.3 & 0.5 &  0.389 & 0.016 & 0.199 & 0.007 \\
  0.266 & 1.527 & 0.017 & 4.920 & 0.017 & 3.016 & 0.021 & 11.1 & 0.4 &  0.344 & 0.013 & 0.230 & 0.006 \\
  1.249 & 1.620 & 0.026 & 4.673 & 0.027 & 2.931 & 0.032 &  9.7 & 0.6 &  0.303 & 0.025 & 0.216 & 0.011 \\
  2.764 & 1.671 & 0.025 & 4.318 & 0.032 & 2.773 & 0.033 &  9.9 & 0.6 &  0.166 & 0.022 & 0.194 & 0.013 \\
  5.846 & 1.755 & 0.030 & 3.860 & 0.037 & 2.428 & 0.039 & 10.2 & 0.6 & -0.041 & 0.022 & 0.230 & 0.024 \\
 13.930 & 1.597 & 0.043 & 3.514 & 0.055 & 2.149 & 0.055 & 14.7 & 0.9 & -0.224 & 0.016 & 0.345 & 0.025 \\
\enddata 
\tablenotetext{a}{The radii in column 1 are measured from 
the center of the galaxy and defined positive towards NW. 
Stellar population parameters (cols. 8, 10, 12) and their uncertainties 
(cols. 9, 11, 13) are derived from the absorption line indices 
(cols. 2--7) using the stellar population models of TMB03.}
\end{deluxetable}

\begin{deluxetable}{rcccrc}
\tablecaption{Observed PNs in NGC 4697 \label{tbl-3}}
\tablewidth{0pt}
\tablehead{
\colhead{Object} & \colhead{Name} & \colhead{RA 2000.0 (hms)} &
\colhead{Decl 2000.0($^o$\thinspace'\thinspace'')} & 
\colhead{AD\tablenotemark{a}} & 
\colhead{m(5007)}} 
\startdata

 1 \ \ \ & W1039   & 12 48 39.015 &  -5 47 53.28 &   47  &   26.0  \\

 2 \ \ \ & W1048   & 12 48 37.846 &  -5 47 38.11 &   38  &   25.7  \\

 3 \ \ \ & W1132   & 12 48 35.902 &  -5 47 25.04 &   38  &   25.5  \\

 4 \ \ \ & W1090   & 12 48 36.655 &  -5 48 42.56 &   42  &   26.1  \\

 5 \ \ \ & W1186   & 12 48 34.791 &  -5 48 35.84 &   37  &   25.7  \\

 6 \ \ \ & W1511   & 12 48 32.967 &  -5 48 22.42 &   48  &   25.6  \\

 7 \ \ \ & W1636   & 12 48 31.260 &  -5 48 30.80 &   75  &   25.7  \\

 8 \ \ \ & W1679   & 12 48 28.651 &  -5 48 06.14 &  108  &   26.6  \\

 9 \ \ \ & W1683   & 12 48 27.123 &  -5 47 58.83 &  131  &   25.6  \\

10 \ \ \ & W1707   & 12 48 27.202 &  -5 49 25.52 &  154  &   25.8  \\

11 \ \ \ & W1709   & 12 48 26.045 &  -5 49 13.11 &  163  &   26.0  \\

12 \ \ \ & W1737   & 12 48 23.645 &  -5 49 58.86 &  217  &   26.0  \\

13 \ \ \ & W1747   & 12 48 20.016 &  -5 49 52.20 &  261  &   26.0  \\

14 \ \ \ & W1746   & 12 48 18.897 &  -5 50 15.67 &  286  &   26.0  \\

\enddata
\tablenotetext{a}{AD is the angular distance from the center of light 
of NGC 4697, measured in arc seconds}
\end{deluxetable}

\begin{deluxetable}{lrrrrrrr}
\tablecaption{Observed relative intensities \label{tbl-4}}
\tablewidth{0pt}
\tablehead{
\colhead{Line} & \colhead{PN 1} & \colhead{PN 2} & \colhead{PN 3} &
\colhead{PN 4} & \colhead{PN 5} & \colhead{PN 6} & \colhead{PN 7}} 
\startdata


 [Ne III]$\lambda$3868   &  1.5 &  1.2 &  2.2 &  2.1 &  1.4 &  1.3 &  2.2 \\


 H$\beta$ $\lambda$4861  &  1.0 &  1.0 &  1.0 &  1.0 &  1.0 &  1.0 &  1.0 \\

 [O III]$\lambda$4959    &  8.0 &  4.2 &  7.2 &  5.1 &  3.1 &  4.3 &  5.9 \\

 [O III]$\lambda$5007    & 23.3 & 14.6 & 22.0 & 15.5 &  9.5 & 12.8 & 18.9 \\

 H$\alpha$ $\lambda$6563 &  3.0 &  3.3 &  3.8 &  3.0 &  3.2 &  3.4 &  4.2 \\

\enddata
\end{deluxetable}

\begin{deluxetable}{lrrrrrrr}
\tablecaption{Observed relative intensities \label{tbl-5}}
\tablewidth{0pt}
\tablehead{
\colhead{Line} & \colhead{PN 8} & \colhead{PN 9} & \colhead{PN 10} &
\colhead{PN 11} & \colhead{PN 12} & \colhead{PN 13} & \colhead{PN 14}} 
\startdata


 [Ne III]$\lambda$3868   &   -  &  2.0 &  1.6 &  2.2 &  1.5 &  2.5 &  1.3 \\


 H$\beta$ $\lambda$4861  &  1.0 &  1.0 &  1.0 &  1.0 &  1.0 &  1.0 &  1.0 \\

 [O III]$\lambda$4959    &  3.5 &  6.0 &  6.2 &  6.2 &  5.0 &  6.7 &  6.1 \\

 [O III]$\lambda$5007    & 10.2 & 18.3 & 19.3 & 19.5 & 14.0 & 18.7 & 18.3 \\

 H$\alpha$ $\lambda$6563 &  3.3 &  4.1 &  3.2 &  3.1 &  2.5 &  3.2 &  2.8 \\

\enddata
\end{deluxetable}

\begin{deluxetable}{rrrrrrrrrrr}
\tablecaption{Illustrative CLOUDY runs\tablenotemark{a} \label{tbl-6}}
\tablewidth{0pt}
\tablehead{
\colhead{logL} & \colhead{HDEN} & \colhead{logT} & \colhead{logRin} &
\colhead{C} & \colhead{N} & \colhead{O} & \colhead{Ne} &
\colhead{5007} & \colhead{3868} & \colhead{L(H$\beta$)}} 
\startdata

 37.43 & 4.60 & 4.98 & 16.3 & 8.50 & 8.00 & 8.60 & 7.88 & 19.0 & 1.9 & 34.99 \\

 37.43 & 4.60 & 4.88 & 16.3 & 8.58 & 8.08 & 8.68 & 7.99 & 19.0 & 1.9 & 34.99 \\

 37.43 & 4.60 & 5.08 & 16.3 & 8.56 & 8.06 & 8.66 & 7.94 & 19.0 & 1.9 & 34.97 \\

 37.43 & 4.40 & 4.98 & 16.3 & 8.52 & 8.02 & 8.62 & 7.91 & 19.0 & 1.9 & 34.99 \\

 37.43 & 4.80 & 4.98 & 16.3 & 8.56 & 8.12 & 8.61 & 7.88 & 19.0 & 1.9 & 35.01 \\

 37.43 & 5.00 & 4.98 & 16.3 & 8.61 & 8.15 & 8.64 & 7.90 & 19.0 & 1.9 & 35.05 \\

 37.43 & 4.60 & 4.98 & 16.1 & 8.58 & 8.15 & 8.61 & 7.90 & 19.0 & 1.9 & 34.92 \\

 37.43 & 4.60 & 4.98 & 16.5 & 8.58 & 8.15 & 8.64 & 7.93 & 19.0 & 1.9 & 35.09 \\

\enddata
\tablenotetext{a}{
The columns give the following quantities: log L of central star 
in erg s$^{-1}$; log nebular H density in cm$^{-3}$; log of the central star 
(blackbody) surface temperature; log of inner nebular radius in cm; C, N, O 
and Ne nebular abundances by number in the usual logarithmic scale with H=12; 
and the log of the H$\beta$ nebular luminosity, in erg s$^{-1}$.
}
\end{deluxetable}

\end{document}